\documentclass[12pt]{article}

\usepackage{scicite}
\usepackage{times}
\usepackage{url}
\usepackage{graphics}
\usepackage{mathtools}
\usepackage{siunitx}
\usepackage{float}
\usepackage{amssymb}

\newenvironment{sciabstract}{%
\begin{quote} \bf}
{\end{quote}}

\title{Observing separate spin and charge Fermi seas in a strongly correlated one-dimensional conductor}

\author
{Pedro~M.~T.~Vianez,${}^{1\ast}$ Yiqing Jin,${}^{1}$ María Moreno,${}^{2}$ Ankita~S.~Anirban,${}^{1}$ \\ Anne Anthore,${}^{3}$ Wooi Kiat Tan,${}^{1}$ Jonathan~P.~Griffiths,${}^{1}$ Ian Farrer,${}^{4}$ \\ David~A.~Ritchie,${}^{1,5}$ Andrew~J.~Schofield,${}^{6}$ \\ Oleksandr Tsyplyatyev,${}^{7\ast}$ Christopher~J.~B.~Ford${}^{1\ast}$\\
\\
\normalsize{${}^{1}$Department of Physics, Cavendish Laboratory, University of Cambridge,}\\
\normalsize{Cambridge, CB3 0HE, UK.}\\
\normalsize{${}^{2}$Departamento de Física Aplicada, Universidad de Salamanca,}\\
\normalsize{Plaza de la Merced s/n, 37008 Salamanca, Spain.}\\
\normalsize{${}^{3}$Universit\'{e} de Paris, C2N, 91120 Palaiseau, France.}\\
\normalsize{${}^{4}$Department of Electronic \& Electrical Engineering, University of Sheffield,}\\
\normalsize{3 Solly Street, Sheffield, S1 4DE, UK.}\\
\normalsize{${}^{5}$Department of Physics, Swansea University,}\\
\normalsize{Vivian Tower, Singleton Park, Swansea, SA2 8PP, UK.}\\
\normalsize{${}^{6}$Department of Physics, Lancaster University, Lancaster, LA1 4YB, UK.}\\
\normalsize{${}^{7}$Institut f\"ur Theoretische Physik, Universit\"at Frankfurt,}\\
\normalsize{Max-von-Laue Stra{\ss}e 1, 60438 Frankfurt, Germany.}\\
\\
\normalsize{$^\ast$To whom correspondence should be addressed; E-mail: pmtv2@cam.ac.uk,}\\
\normalsize{o.tsyplyatyev@gmail.com, cjbf@cam.ac.uk}
}

\date{}

\begin{document}

\baselineskip24pt

\maketitle

\newpage
\begin{sciabstract}
An electron is usually considered to have only one form of kinetic energy, but could it have more, for its spin and charge, by exciting other electrons? In one dimension (1D), the physics of interacting electrons is captured well at low energies by the Tomonaga-Luttinger model, yet little has been observed experimentally beyond this linear regime. Here, we report on measurements of many-body modes in 1D gated-wires using tunnelling spectroscopy. We observe two parabolic dispersions, indicative of separate Fermi seas at high energies, associated with spin and charge excitations, together with the emergence of two additional 1D `replica’ modes that strengthen with decreasing wire length. The effective interaction strength is varied by changing the amount of 1D inter-subband screening by over 45\%. Our findings demonstrate the existence of spin-charge separation in the whole energy band outside the low-energy limit of validity of the Tomonaga-Luttinger model, and also set a constraint on the validity of the newer nonlinear Tomonaga-Luttinger theory.
\end{sciabstract}


\section*{Introduction}

Many-body systems cannot be explained by studying their individual components, with interactions often giving rise to collective excitations from which an array of qualitatively new quasi-particles starts to emerge. This is particularly striking in one dimension (1D), as here geometrical confinement alone imposes strong correlations in the presence of any interactions, leading to well-known non-Fermi-liquid phenomena such as spin-charge separation \cite{Giamarchi_book}. Overall, the behaviour of 1D interacting gapless systems in the low-energy regime is well captured by the Tomonaga-Luttinger model \cite{Tomonaga50,Luttinger63,Haldane81}, and has been extensively tested in carbon nanotubes \cite{Bockrath99,ishii03,Shi15}, semiconductor quantum wires \cite{Auslaender05,Jompol09,laroche_positive_2011}, antiferromagnets \cite{Kim06} and more recently, cold-atom chains \cite{Vijayan20}. The model, which assumes a linearised single-particle dispersion, is expected to only be valid close to the Fermi points, where nonlinearities are still weak. However, pronounced consequences of band curvature have also very recently started to be explored experimentally \cite{Barak10,Jin19,Wang20}. 

At the same time, modelling such systems is a long-standing open problem. Simultaneous introduction of the charge and spin degrees of freedom into a  nonlinear extension of Tomonaga-Luttinger-liquid (TLL) theory \cite{Imambekov09} predicted that the spin-charge separation  would no longer exist beyond the low-energy regime  \cite{Schmidt10,Schmidt10p}, since the holons (i.e., charge-type excitations) are made unstable by the nonlinearities. Instead, a mixture of spinons (i.e., spin-type excitations) and holons is responsible for the power-law threshold behaviour around the spectral edges.  Extra, higher-order 1D modes, which have the spectral edge dispersion shifted and mirrored from that of the main 1D subband, were predicted in \cite{Imambekov09} as well.
Another theory \cite{Tsyplyatyev15} showed that these extra modes should only emerge as the system length is reduced.
We have observed some signatures of these `replicas' in the past \cite{Tsyplyatyev16,Moreno16}, but experimental results beyond the linear regime that have both enough resolution and clarity to distinguish between predictions have been lacking.

Here, we measure the spectral function for the spin and charge excitations well beyond the linear regime using a tunnelling spectroscopy technique that allows mapping in both energy and momentum space. In the amplitude of our signal, we observe  for the first time how the two branches of the linear TLL modes evolve away from the Fermi points into two fully formed nonlinear dispersions that consist of purely spin or charge collective modes, identified by comparison with the spectra predicted by the 1D Fermi-Hubbard model \cite{LiebWu68}. Both dispersions are parabolic in shape but with different masses, implying the existence of two Fermi seas of different types. This result shows that the spin and charge collective excitations both remain stable in the whole conduction band, well beyond the low-energy limit of the original Tomonaga-Luttinger model where their existence was first established. We are able to tune the degree of screening of the Coulomb interaction by changing the confinement in our wires and so the number of occupied subbands. This is accompanied by a variation of approximately 45\% of the two-body interaction energy and allows us to trace how both Fermi seas evolve as the interaction strength is varied. Measuring wires of different lengths, we are also able to observe two, with a possible third, separate 1D nonlinear `replica' modes of the spinon type that systematically emerge as the length decreases.

\begin{figure*}
  \includegraphics[width=0.99\textwidth]{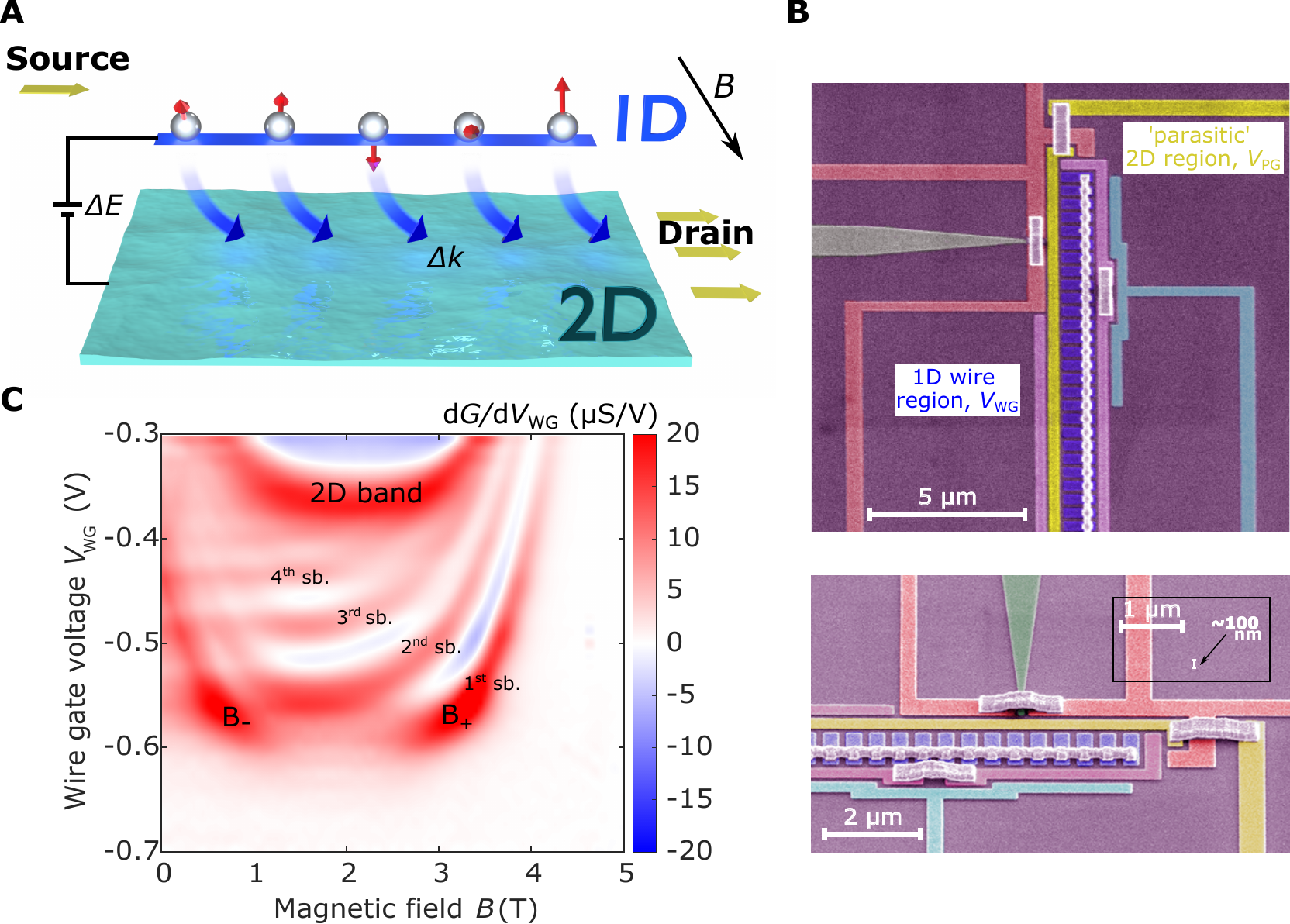}
  \caption{\small \textbf{Mapping a 1D system via magnetotunnelling spectroscopy.} (\textbf{A}) Schematic representation of the 1D-2D spectrometer device. We measure momentum-resolved tunnelling to and from an array of 1D wires (only one wire shown here for simplicity) and a 2D electron system, and map the elementary excitations in each system by measuring the tunnelling conductance while varying both their energy $\Delta E \propto V_{\textrm{DC}}$ and momentum $\Delta k \propto B$. Current flows from the source into the wire and tunnels between the layers in order to reach the drain. (\textbf{B}) Scanning electron micrographs of the various surface gates present in our device. See Materials and Methods for details on gate operation and how to set up the tunnelling regime. Inset: Air-bridge interconnections between surface gates. (\textbf{C}) 1D wire subbands participating in the tunnelling process. We observe from four to one 1D subbands before the wires pinch off.}
  \label{fig:fig1}
\end{figure*}

\section*{Results}
\subsection*{Characterisation of the 1D array}
Our experiment consists of a tunnelling spectrometer made on a $\textrm{GaAs}/\textrm{Al}_{0.33}\allowbreak\textrm{Ga}_{0.67}\textrm{As}$ heterostructure with two parallel quantum wells (QWs), grown by molecular-beam epitaxy (MBE). We measure the tunnelling conductance $G={\textrm{d}}I/{\textrm{d}}V_{\textrm{DC}}$ between the 1D wires and the 2D layer at lattice temperatures $T\sim300$\,mK (see Fig.\ \ref{fig:fig1}A), where $I$ is the tunnelling current while $V_{\textrm{DC}}$ the DC bias applied between the layers. Tunnelling occurs when filled states in one system have the same energy and momentum as empty states in the other, therefore ensuring that both energy and momentum are conserved. In order to map the dispersion of each system, a negative (positive) voltage $V_{\textrm{DC}}$ applied to the 1D wires provides energy for tunnelling from (to) 1D states below (above) the Fermi level, while an in-plane magnetic field $B$ perpendicular to the wires boosts the momentum, offsetting the spectral functions of each system by $\Delta k=eBd/\hbar$, where $e$ is the electronic charge and $d$ the separation between the wells. The differential tunnelling conductance $G$ displays resonant peaks corresponding to maximal overlap of the offset spectral functions (dispersion relations). The device therefore behaves as a spectrometer, with the well-characterised 2D system being used to probe the less-well understood spectral function of the 1D system. 

We use a surface-gate depletion technique in order to establish separate contacts to each well. Our 1D system consists of an array of $\sim$ 400 highly regular quantum wires formed in the upper layer by using a set of wire gates (WGs) fabricated on a Hall bar via standard electron-beam lithography and connected by air bridges (see Fig.\ \ref{fig:fig1}B and inset) \cite{jin_microscopic_2021}. Use of an array averages out impurities, length resonances and charging effects as well as increasing the overall strength of the measured signal. For the shorter devices, the air bridges are crucial for ensuring that good uniformity is obtained along the entire length of the wire, which would otherwise become narrower at one end if instead all the gates were joined by a continuous metal strip. Current is injected into the 1D wires via a small region, 0.45\,$\mu$m wide. Unlike the wires, however, this region is 2D in nature and its parasitic signal can be readily distinguished from the 1D signal in the measured data since its density is different. We use the unconfined weakly interacting 2D electron gas (2DEG) in the bottom well as a well-understood spectrometer. Note that our measurement is subject to capacitive effects between the wells, as well as between each well and the surface gates, and we take these into account in all the curves we plot, see Supplementary Text section 1.6.1.

\begin{figure*}
  \includegraphics[width=0.99\textwidth]{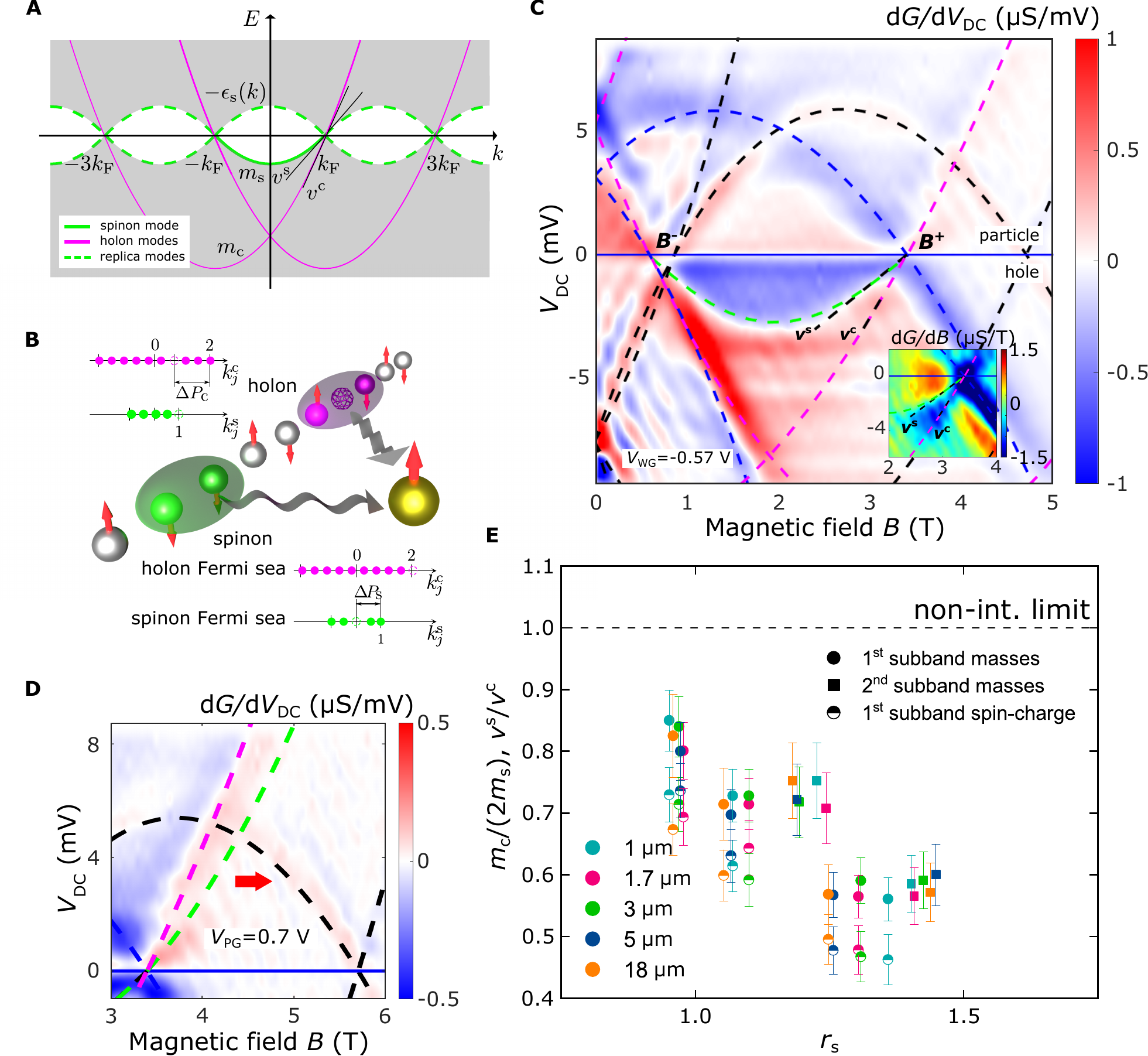}
  \caption{\small \textbf{Two Fermi seas.} 
  (\textbf{A}) Dispersion proposed for an interacting 1D system (grey: continuum of many-body excitations, green lines: spinon modes, magenta lines: holon modes, dashed: replicas). (\textbf{B}) Graphical representation of the decomposition of an electron into a spinon and a holon in a tunnelling process, along with an illustration of two distinct Fermi seas (filled dots) describing the pure holon (top) and spinon (bottom) excitations (in units of $k_{\textrm F}$)---see text. (\textbf C) Map of the tunnelling conductance ($G$) differential ${\textrm{d}}G/{\textrm{d}}V_{\textrm{DC}}$ \textit{vs} DC bias $V_{\textrm{DC}}$ and in-plane magnetic field $B$, for a $5\,\mu$m-long device. Superimposed curves mark all possible single-electron tunnelling processes, together with the resonant dispersions of the spin and charge modes marked by dashed green and magenta lines, respectively. Inset: ${\textrm{d}}G/{\textrm{d}}B$ around the $+k_{\textrm{F}}$ point at negative biases where both spin ($v^{\textrm{s}}$) and charge ($v^{\textrm{c}}$) lines can be seen. (\textbf{D}) ${\textrm{d}}G/{\textrm{d}}V_{\textrm{DC}}$ above $+k_{\textrm{F}}$ ($\sim 3.3$\,T), where $V_\textrm{PG}=0.7$\,V so that the parasitic signal has moved further to the right.  We systematically observe no signal from the spinon mode at $V_\textrm{DC}>0$ in all devices that we measured, see text for discussion. A na\"ive extension of the spinon dispersion, observed in the hole sector, into the particle sector is given by the green-dashed line. (\textbf{E}) Ratio of holon-to-spinon masses and spinon-to-holon velocities \textit{vs} interaction parameter $r_{\textrm{s}}$ for devices of different lengths.}
  \label{fig:fig2}
\end{figure*}

A plot of ${\textrm{d}}G/{\textrm{d}}V_\textrm{WG}$ vs $B$ and $V_{\textrm{WG}}$ shows U-shaped curves, one per 1D subband (Fig.\ \ref{fig:fig1}C). We start our experiment by choosing $V_{\textrm{WG}}$ so that there is just one 1D subband occupied. Fig.\ \ref{fig:fig2}C shows an example of such a measurement, with conductance through the sample being measured as a function of energy ($\propto V_{\textrm{DC}}$) and momentum ($\propto B$). Here, the tunnelling map can be divided into two sectors, particle (for $V_\textrm{DC}>$0) and hole (for $V_\textrm{DC}<$0), corresponding, respectively, to electrons tunnelling into and out of the wires. The 1D Fermi wave vector $k_{\textrm{1D}}=ed(B^{+}-B^{-})/2\hbar$ is determined from the crossing points ($B^{-}$ and $B^{+}$) along the $V_{\textrm{DC}}=0$ line. The electron density in the wires is $n_{\textrm 1D}=2k_{\textrm{1D}}/\pi$, which then gives the interaction parameter $r_{\textrm{s}}= 1/\left(2a_{\textrm{B}}^\prime n_{\textrm{1D}}\right)$, where $a_{\textrm{B}}^\prime$ is the Bohr radius of conduction electrons in GaAs, see Supplementary Text section 1.5. The density can be controlled by tuning $V_{\textrm{WG}}$, reducing it down to $n_{\textrm{1D}}\sim30\,\mu $m$^{-1}$ before the wires pinch off.

\subsection*{Observation of two Fermi seas for spin and charge}
The curves drawn over the data in Fig.\ \ref{fig:fig2}C are those expected from single-electron tunnelling processes. Undesirable yet unavoidable `parasitic' tunnelling coming from the narrow 2D injection region (marked by the dashed black curves, see Supplementary Text section 1.6.2) produces a background in the form of a set of parabolic dispersions, which can be subtracted once separately mapped (with the wire gates pinched off). On the other hand, dashed blue curves reveal the elementary excitations of the 2D lower well, as probed by the 1D wires. The remaining strong features, marked by the green and magenta dashed curves, arise from the 1D system. We are unable to explain these data well using a single parabola (see Supplementary Text section 1.6.3 for details), a fact that strongly points in the direction of separate spin and charge modes. 

In order to identify whether this is the case, we interpret the 1D tunnelling signal using the dispersion of the 1D Fermi-Hubbard model in the semiconductor limit, in which many-body spectra are described completely by the Lieb-Wu equations \cite{LiebWu68}. This system of nonlinear coupled equations is solved for two types of momentum states, $k^{\textrm{c}}_j$ (for charge) and $k^{\textrm{s}}_j$ (for spin degrees of freedom), which for the ground state form two filled Fermi seas marked by filled circles of two different colours in Fig.\ \ref{fig:fig2}B (for a detailed discussion of the theoretical model, see sections 1.1, 1.2, and 1.3 of the Supplementary Text). An excitation, say an electron tunnelling out of the wire, removes one charge and one spin simultaneously to reassemble a free electron, marked by a pair of green and magenta empty circles in Fig.\ \ref{fig:fig2}B. Placing the hole of one type at its corresponding Fermi energy and moving the other one through the band describes the spectrum of the purely holon or purely spinon modes. While the momentum of these collective excitations as a whole is well-defined due to the translational invariance, with $k=k_{\textrm{F}}-\Delta P_{\textrm{c}}$ or $k=k_{\textrm{F}}-\Delta P_{\textrm{s}}$, the constituent degrees of freedom form non-equidistant distributions of their (quasi-)momenta, which depend in detail on the interaction strength and the positions of the two holes, owing to the strongly correlated nature of the model. Explicit solution of the Lieb-Wu equations for $k^{\textrm{c}}_j$ and $k^{\textrm{s}}_j$ for the two kinds of pure excitations produces the two dispersions drawn as the magenta and green solid lines in Fig.\ \ref{fig:fig2}A. These two dispersions constructed out of collective modes for spin and charge have a shape close to parabolic, and have a simple description in terms of two Fermi seas with different masses. This closely matches our experimental observations, implying  the presence of these two Fermi seas (see section 1.6 of the Supplementary Text) in a real system of interacting electrons.

Around the $\pm k_{\textrm F}$ points, these curves are almost linear, and can be characterised by two different slopes $v^{\textrm{c}}$ and $v^{\textrm{s}}$, parameters of the spinful Tomonaga-Luttinger model for the holon and spinon modes, respectively. These two velocities are related microscopically to the Hubbard interaction parameter $U$ \cite{Schultz90,Frahm90}, and the spectral function predicted by the linear Tomonaga-Luttinger theory
displays two strong peaks on these two branches \cite{Schoenhammer92,Voit93}, which have already been measured in semiconductor quantum wires \cite{Auslaender05,Jompol09}.
Away from the Fermi points, the spectra of holons and spinons
extend naturally to the nonlinear region, evolving into two separate curves that are close to parabolae described by masses $m_{\textrm{c}}$ and $m_{\textrm{s}}$, respectively. These shapes indicate formation of two separate Fermi seas by the nonlinear excitations. Their dispersions cross the Fermi energy at two different pairs of Fermi points ($\pm k_{\textrm{F}}$ and $\pm 3k_{\textrm{F}}$), see a numerical simulation of the Fermi-Hubbard model via  the dynamical
density-matrix renormalization group method in \cite{benthien_spectral_2004}, since the number of holons is twice the number of spinons for the spin-unpolarised wires in our experiments, making the densities for the two kinds also different by the same factor. The ratio of their masses depends on $U$, deviating further from the free-particle value $m_{\textrm{c}}/(2m_{\textrm{s}})=1$ with increasing interaction strength. 

As we have seen, the dispersion of the strongest features in the experimental 1D signal (marked by dashed green and magenta lines in Fig.\ \ref{fig:fig2}C) cannot simply be interpreted using only a single parabola, corresponding to a single Fermi sea. This can be further established by analysing the tunnelling signal at zero field, see Supplementary Text section 1.6.4 for details. If, instead of just one parabola, we use two, corresponding to two Fermi seas, such as that predicted by the Fermi-Hubbard model, then we can match the experiments well. We observe two modes in our data, which match the dispersions of pure excitations of the two different kinds depicted in Fig.\ \ref{fig:fig2}A, where two distinct Fermi seas are formed by the nonlinear spinon and holon collective modes out of the many-body continuum away from the Fermi points. This result demonstrates a non-perturbative  effect that interactions cause in the whole band in 1D, posing a new theoretical challenge of accounting for higher-order processes beyond what has been considered in the literature so far \cite{Mahan_book} to describe it.

While we observe two dispersions in the hole sector, only one is visible in the particle sector, which we easily associate with the holon Fermi sea, as its gradient matches that of the charge line in the hole sector. At the same time, the spinon dispersion that assumes a na\"{i}ve extension from the hole sector is systematically absent in the particle sector for all devices measured, see green dashed line in Fig.\ \ref{fig:fig2}D. This result is, however, compatible with the particle-hole asymmetry in relaxation times of hot carriers as reported previously in \cite{Barak10}. Even though spin and charge excitations were not resolved in \cite{Barak10}, assuming that the spinon branch in the particle sector is unstable (so that we do not observe it), there would be an accelerated relaxation for hot electrons as they eventually split into spinon and holons.
Note that the spinon Fermi sea alone has already been observed by neutron scattering (probing the dynamic structure factor instead of the spectral function that we measure here) in antiferromagnetic spin chains realised in insulating materials \cite{Lake05,Mourigal13,Lake13} as a spectral edge with a nonlinear dispersion separating the multi-spinon continuum from a forbidden region \cite{Caux05,Goehmann04}. In these experiments the spectral power of the excitations drops very rapidly towards the particle part of the spinon dispersion making it therefore undetectable. Also note that, in the present experiment, the charges are delocalised as well, permitting us to see both Fermi seas at the same time.

By tuning the confinement in the wires we are also able to change the number of occupied subbands and their respective densities, therefore allowing us to follow the evolution of each Fermi sea as $r_{\textrm{s}}$ is changed by a significant amount. Such statistics collected from a range of samples in Fig.\ \ref{fig:fig2}E show a systematic trend of larger deviations of the observed $m_{\textrm{c}}/(2m_{\textrm{s}})$ ratio from its non-interacting value with increasing $r_{\textrm{s}}$.  The ratio of the Luttinger parameters $v^{\textrm{s}}/v^{\textrm{c}}$ simultaneously extracted from the same data (see as an example the inset in Fig.\ \ref{fig:fig2}C) exhibits a very similar dependence on $r_{\textrm{s}}$.  

\subsection*{Further interaction signatures: a hierarchy of 1D `replica' modes} 

\begin{figure*}
  \includegraphics[width=0.99\textwidth]{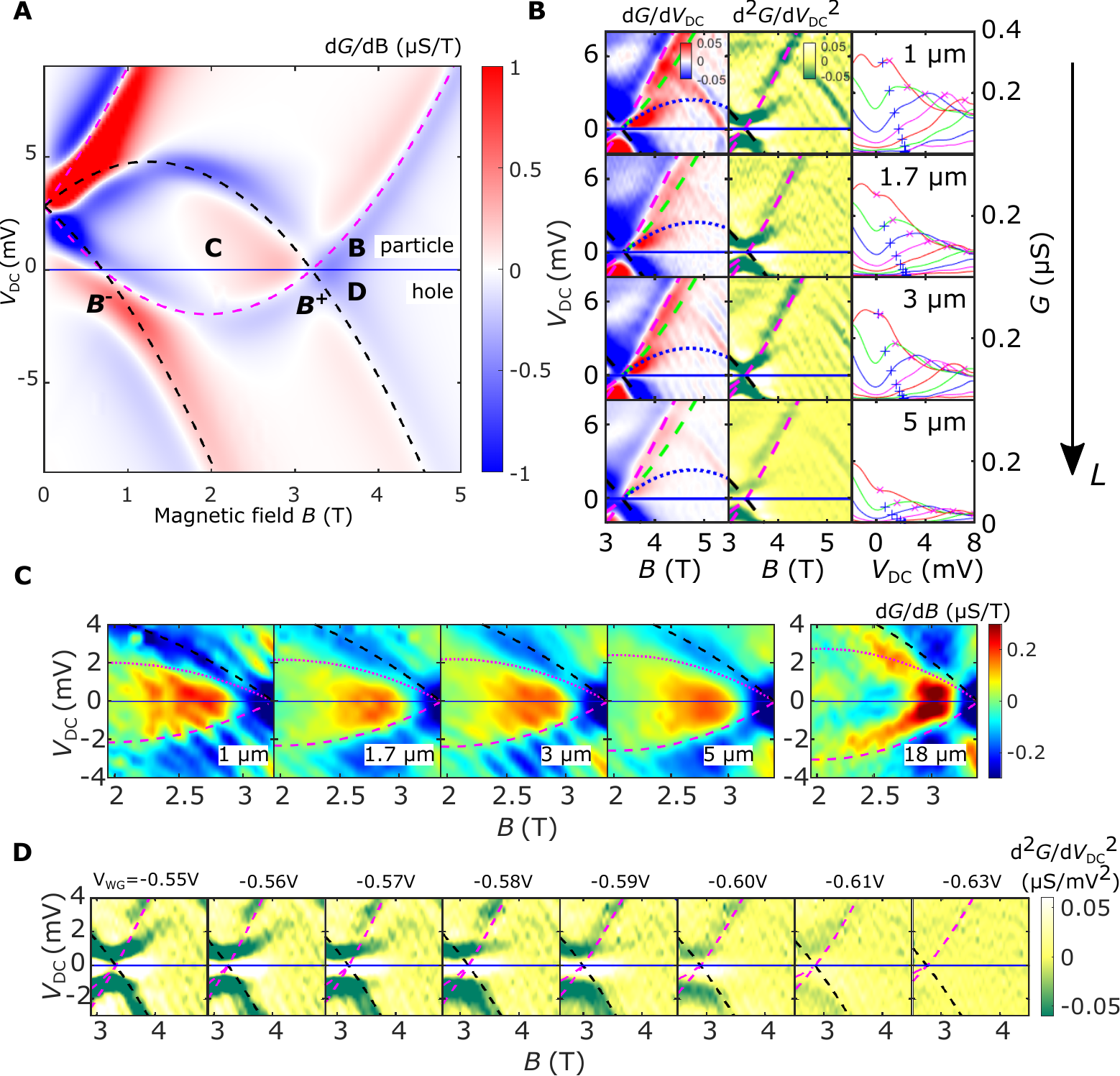}
  \caption{\small \textbf{A Hierarchy of Modes.} (\textbf{A}) Simulated map of the differential tunnelling conductance ${\textrm{d}}G/{\textrm{d}}B$ vs $V_{\textrm{DC}}$ and $B$, between a 1D non-interacting system (magenta) and a 2DEG (black). In the absence of interactions the spinon and holon dispersions are degenerate with each other. (\textbf{B}) ${\textrm{d}}G/{\textrm{d}}V_{\textrm{DC}}$ (left) and ${\textrm{d}}^2G/{\textrm{d}}V_{\textrm{DC}}^2$ (centre), for devices of different lengths, as labelled. Right column: $G$ vs $V_{\textrm{DC}}$ at various fields $B>B^{+}$ for the data in the matching plots to the left; `x' and `+' symbols on each curve indicate the position of the fitted dispersions in the particle sector for the holon branch and the first-order `replica', respectively (see text for definition)---$G$ stays high between the two. (\textbf{C}) ${\textrm{d}}G/{\textrm{d}}B$ for $B<B^{+}$, showing another first-order `replica'  mode (dotted magenta) in the particle sector for a variety of different-length devices. (\textbf{D}) ${\textrm{d}}^2G/{\textrm{d}}V_{\textrm{DC}}^2$ for a $1\,\mu$m device at a variety of different wire-gate voltages $V_{\textrm{WG}}$. This `replica' mode responds to changes in $V_{\textrm{WG}}$, completely disappearing once the wires are pinched off. Symmetric to it, in the hole sector, a kink in conductance can be observed, only visible in our shortest $1\,\mu$m devices. Conductance has been normalised by device length in \textbf{B}, \textbf{C} and \textbf{D}.}
  \label{fig:fig4}
\end{figure*}

Having now identified two separate Fermi seas for spin and charge in our data, we further analyse the 1D dispersion by contrasting it with the simulated tunnelling conductance map between a non-interacting 1D system and a 2DEG (see Fig.\ \ref{fig:fig4}A). Note how, unlike in Fig.\ \ref{fig:fig2}C, it is possible to fit both particle and hole sectors of the map with a single parabola (dashed magenta). This is because, in the absence of interactions, the opposite spin states are degenerate, leaving room for only a single Fermi sea. We start by examining the region just above $B^{+}$ (i.e. +$k_{\textrm{F}}$), where a clear feature not accounted for by our non-interacting simulation can be observed, see Fig.\ \ref{fig:fig4}B. Here, the tunnelling conductance peak broadens and splits, with one boundary following the 1D holon mode while the other branches away from it. In order to isolate it from any potential background contamination, we apply a positive $V_\textrm{PG}$ again in order to move the `parasitic' resonance signatures away from $B^{+}$. We also observe that this extra feature is not visible once the wires are past pinch-off, and that it is independent of the `parasitic' tunnelling signal. 

The mode-hierarchy picture for fermions \cite{Tsyplyatyev15,Tsyplyatyev16} predicts that the continuum of the many-body excitations separates itself into levels (i.e., first and higher orders) by their spectral strengths, which are proportional to integer powers of $R^2/L^2$ (i.e., this parameter to the first and higher powers), where $R$ is the length scale related to the interaction and $L$ is the length of the system (see Supplementary Text section 1.4 for more details). The principal (spin and charge) parabolae have then the largest amplitude, while their mirrors with respect to the chemical potential (\emph{i.e.}, $V_\textrm{DC}=0$) and translations by integer multiples of $2k_\textrm{F}$ manifest themselves as `replicas', see dashed lines on Fig.\ \ref{fig:fig2}A, with amplitudes proportional to higher powers of the small parameter $R^2/L^2$. We have observed this feature in samples with wire lengths ranging from 1\,$\mu$m to 18\,$\mu$m, with all devices being mapped at very similar densities and Fermi energies, making them otherwise similar in $R$. In all of them the strength of the mode marked by the dotted blue line in Fig.\ \ref{fig:fig4}B, which is a `replica' of the parabola formed by nonlinear spinons, decreases as the $B$ field is increased away from the crossing point. However, once the background has been subtracted and $G$ has been normalised by length, one can see qualitatively that the decay away from $B_{+}$ was slower the shorter the 1D system, as predicted, with the signal vanishing at higher momenta away from $+k_{\textrm{F}}$.

In order to test our prediction of length-dependent spin-type `replica' modes further, we have looked at two other sectors of the tunnelling map, see Fig.\ \ref{fig:fig4}C and D. We initially reported the first mode between $\pm k_{\textrm{F}}$ as an inverted (spinon) shadow band symmetric to the 1D (spinon) mode in \cite{Moreno16}. According to the nonlinear theory of TLLs \cite{Imambekov09,Glazman12}, in the main $|k|<k_{\textrm{F}}$ region of the hole sector, the edge of support (defined as the hole excitation with the smallest possible energy for a given momentum) is predicted \cite{Schmidt10,Essler10} to coincide with the spinon mass shell, whose dispersion $\epsilon_{\textrm{s}}(k)$ we have already observed to be very close to a parabola in our experiment. Similarly, in the main region of the particle sector, the edge of support is also predicted to be given by the inverted spinon mass shell $-\epsilon_{\textrm{s}}(k)$ in Fig.\ \ref{fig:fig2}A. Consistent with the nonlinear theory, a symmetric inverted replica was seen in the particle sector, opposite to the main 1D subband, in all mapped devices, up to 5$\,\mu$m (see Fig.\ \ref{fig:fig4}C). This feature can also be seen in Fig.\ \ref{fig:fig2}C. According to the mode-hierarchy picture, a length dependence similar to that of Fig.\ \ref{fig:fig4}B is also expected to be observed since this is also a sub-leading mode. Although such dependence is not particularly clear from $1-5\,\mu$m, the replica mode was seen to not be present at all for the 18\,$\mu$m wire.

Similarly, another `replica' mode is also predicted to exist at $k_{\textrm{F}}<k<3k_{\textrm{F}}$, symmetric to the sub-leading spinon mode shown in Fig.\ \ref{fig:fig4}B, but in the hole sector. Only for the shortest, 1\,$\mu$m device is a feature consistent with this picture starting to be observed, hinting that a full observation of this mode would probably only happen at sub-micron lengths. Nevertheless, as seen in Fig.\ \ref{fig:fig4}D, both modes evolve in tandem with each other as the 1D channels are squeezed towards pinch-off. This further establishes that these features are 1D in nature and cannot originate from the `parasitic' injection region. All three replica features discussed emerge as the effective length of the 1D system is reduced, compatible with the mode hierarchy picture where a level hierarchy emerges controlled by system's length. We attribute the different lengths at which they become visible in this experiment (the first `replica' only being observed below $5\,\mu$m, the second below $18\,\mu$m and the third `replica' only at $1\,\mu$m) to different numerical prefactors that are still unknown theoretically for spinful systems. Nevertheless, the fact that we are observing features compatible with the mode-hierarchy picture further establishes our technique as being capable of detecting interaction effects in the nonlinear regime.

\subsection*{Evolution of the two Fermi seas with interaction strength}

\begin{figure*}
  \centering
  \includegraphics[width=0.99\textwidth]{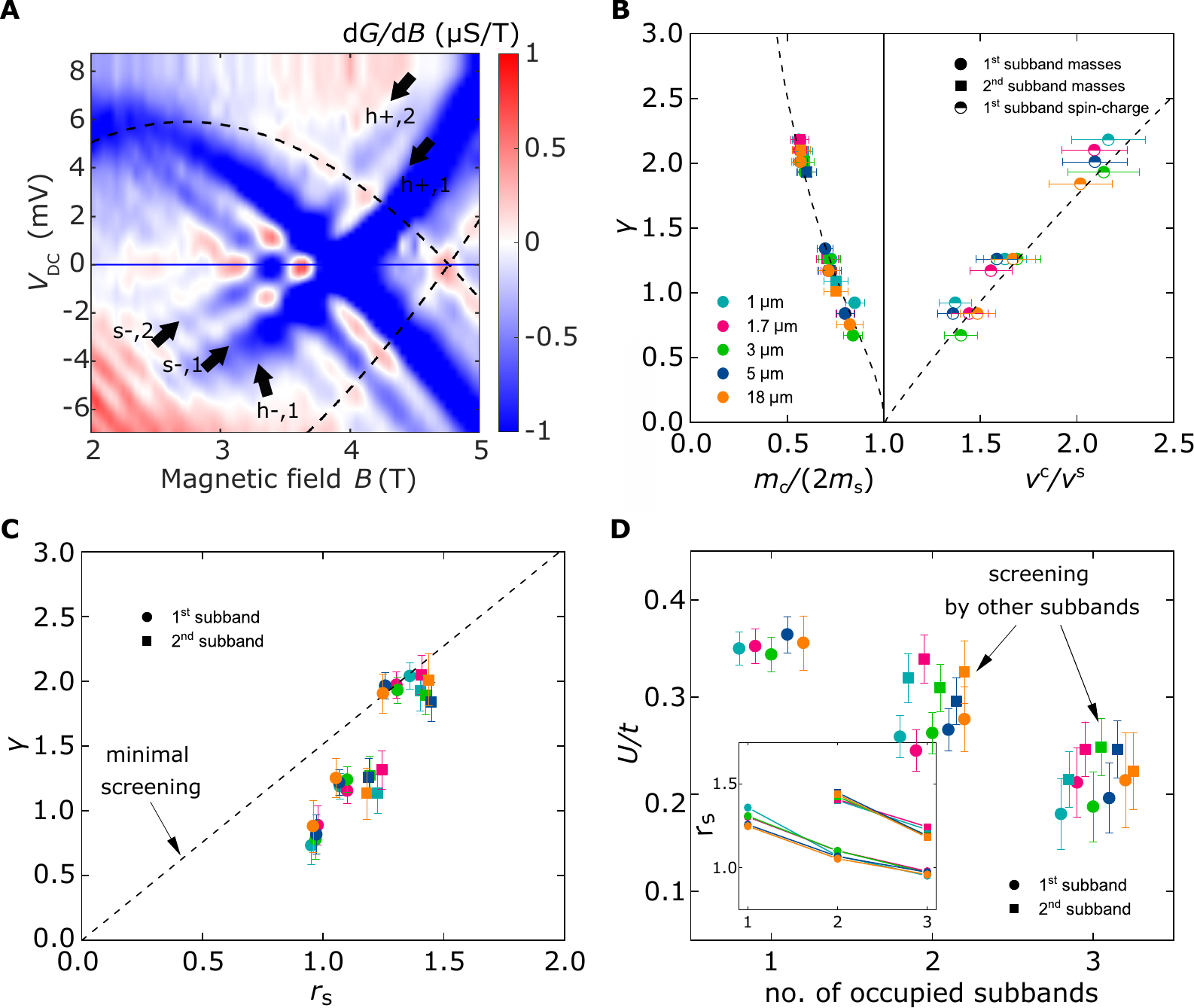}
  \caption{\small \textbf{1D-1D screening.} (\textbf{A}) ${\textrm{d}}G/{\textrm{d}}B$ of a $5\,\mu$m device mapped in the multi-subband regime. Dashed black line marks the location of the subtracted 2D-2D `parasitic' signal. Arrows point to the location of spinon (s) and holon (h) modes, in both the hole (-) and particle (+) sectors, for each occupied subband (1, 2). (\textbf{B}) Macroscopic dimensionless Hubbard-model parameter $\gamma$ (see text for discussion) \textit{vs} mass and velocity ratio, for devices of different lengths, as extracted by fitting the bottom two subbands. The Hubbard model (dashed black) can reproduce well the observed experimental dependence. (\textbf{C}) $\gamma$ vs the interaction parameter $r_{\textrm{s}}$, where an approximate linear dependence can be seen. The dashed curve corresponds to fitting using only data from the single-subband-occupancy regime. Note that, even when allowing for errors, all remaining points fall systematically below this line, indicating the presence of 1D-1D inter-subband screening. (\textbf{D}) Hubbard parameter $U/t$ \textit{vs} number of occupied subbands as extracted from $\gamma$. The asymmetry in screening between the first and second subbands is expected from their difference in densities. Points slightly offset horizontally from each other for clarity. Inset: Interaction parameter $r_{\textrm{s}}$ \textit{vs} number of occupied subbands. We can change $r_{\textrm{s}}$ by tuning the 1D density $n_{\textrm{1D}}$ of the wires (see Fig.\ S4B in Supplementary Materials).}
  \label{fig:fig5}
\end{figure*}

Up to now we have confined our analysis to dispersion maps in the single-subband regime. In the current geometry, however, we are also able to vary the number of occupied subbands up to four, by tuning the wire-gate voltage $V_{\textrm{WG}}$ until the upper layer starts to become 2D when carriers delocalise between the wires. While the emergent hierarchy of modes becomes almost impossible to see in the data with more than one subband occupied, the parameters of two Fermi seas can still be quite reliably extracted, see Fig.\ \ref{fig:fig5}A and Supplementary Text section 1.6.5. Variation of the number of subbands provides us with an additional tool for assessing the microscopic interaction parameter of the Hubbard parameter $U$ in our experiment at the quantitative level. 
The macroscopic dimensionless parameter controlling the Hubbard model in 1D is \cite{OT14}
\begin{equation}
\gamma=0.032\frac{\lambda_{\textrm{F}}}{a}\frac{U}{t},\label{eq:gamma_U}
\end{equation} 
where $a$ is the lattice parameter of the underlying crystal, $t$ the hopping amplitude, and $\lambda_\textrm{F}$ the Fermi wavelength. For $\gamma<1$, the weakly interacting electrons are almost spin-degenerate, having double occupancy for each momentum state, as for free particles. For $\gamma>1$, each momentum state is occupied by only one electron due to strong Coulomb repulsion. Such a dependence of the system's behaviour on $\gamma$ is qualitatively the same as the dependence on $r_{\textrm{s}}$ in all dimensions, reflecting the ratio of the total interaction energy to the kinetic energy.

We have varied the number of occupied subbands all the way up to four, and fitted the bottom two using the model of two Fermi seas to extract the ratio of masses and velocities. Fitting the same data with the dispersion produced by the Hubbard model in a similar fashion to what was done before for the single-subband regime, see Fig.\ \ref{fig:fig2}C, we obtain the values of $\gamma$ that correspond to these ratios for each individual subband and for each subband occupancy in Fig.\ \ref{fig:fig5}B, which allows the data points from multiple wires with a variety of densities (as seen in Fig.\,2E) to be collapsed on to the same curve. The agreement between experiment and theory seen here further reinforces our claim of two separate Fermi seas for spinons and holons, having collected statistics belonging to five different devices, fabricated from two different wafers, across different fabrication cycles, and measured independently from one another in different cool-downs. Comparing the already extracted values of $r_{\textrm{s}}$ with $\gamma$ for all measurements in Fig.\ \ref{fig:fig5}C, we find that the two are approximately proportional to each other with a coefficient of $\simeq1.5$. We interpret the still-observable discrepancy as a manifestation of the screening effect that is not captured by $r_{\textrm{s}}$ but is taken into account explicitly in the Hubbard model via the two-body interaction energy $U$. The latter is proportional to the integral of the screened Coulomb potential, in which only the screening radius is changed in our experiment. 

By means of the relation in Eq.\ (\ref{eq:gamma_U}) we extract the evolution of $U/t$ from the already obtained values of $\gamma$ and $\lambda_\textrm{F}$ as a function of the number of occupied subbands, as shown in Fig.~\ref{fig:fig5}D, for both the 1$^{\textrm{st}}$ and 2$^{\textrm{nd}}$ lowest subbands, in different-length systems. Data corresponding to four occupied subbands was excluded, as its proximity to the non-interacting limit made the fitting less reliable. Similarly, fitting to the dispersions of the third and fourth subbands was not attempted owing to the lack of sharp features and overall increase in map complexity. Nevertheless, two clear trends emerge: first, $U$ decreases as more subbands are progressively filled, resulting in relative reductions of about $\sim45$\% for the 1$^{\textrm{st}}$ subband and of $\sim25$\% for the $2^{\textrm{nd}}$ subband; second, the bottom subband seems to be systematically more strongly screened than the second, most likely due to their difference in density. These trends are also apparent in Fig.\ \ref{fig:fig5}C, where the dashed line is drawn through the single-subband points and all further occupied subband systematically fall below it. We note that the values measured in the single-subband regime are consistent with the estimates made in \cite{Glazman_quantum_1992}, where the long-range Coulomb interaction between the electrons is screened by a conducting plate a certain distance away from the 1D wire. As far as the authors are aware, this is the first clear observation of screening effects between two 1D systems, with similar conclusions reported by \cite{Kim20} in 2D systems. 

\section*{Discussion}
We have shown that spin-charge separation is more robust than previously thought, extending past the low-energy regime of the TLL to beyond the Fermi energy. By tuning the degree of screening of the Coulomb interaction by changing the confinement in our wires and so the number of occupied subbands, we saw how the masses associated with the spin and charge Fermi seas evolve as a function of the interaction strength, with a remarkably good quantitative agreement with the predictions of the 1D Fermi-Hubbard model. At the same time, our comparison of quantum wires of different lengths confirms the prediction of the mode-hierarchy theory, observing systematically the emergence of at least two `replica' modes as the wire length decreases.

\section*{Materials and Methods}
\subsection*{Device fabrication}

All tunnelling devices measured in this work were fabricated using double-quantum-well heterostructures, grown via molecular-beam epitaxy (MBE), comprised of two identical 18\,nm GaAs quantum wells separated by a 14\,nm Al$_{0.165}$Ga$_{0.835}$As superlattice tunnel barrier [10 pairs of Al$_{0.33}$Ga$_{0.67}$As and GaAs monolayers]. On each side of the barrier there were 40\,nm Si-doped layers of Al$_{0.33}$Ga$_{0.67}$As (donor concentration $1\times 10^{24}\textrm{m}^{-3}$), with the lower and upper spacers being respectively 40\,nm and 20\,nm wide. Wafer 1, however, differed from Wafer 2 by having an additional 100$\times$2.5\,mm/2.5\,mm GaAs/AlGaAs superlattice below the 350\,nm AlGaAs under the lower quantum well. This resulted in electron concentrations of about 3 (2.2) $\times 10^{15}\textrm{m}^{-2}$ with mobilities of around 120 (165) $\rm{m^{2}V^{-1}s^{-1}}$ in the top (bottom) well of Wafer 1, while 2.85 (1.54) $\times 10^{15}\textrm{m}^{-2}$ and 191 (55) $\rm{m^{2}V^{-1}s^{-1}}$ for Wafer 2, as measured at 1.4\,K. A 10\,nm GaAs cap layer was used to prevent oxidation. The distance from the upper well to the surface was $\sim$ 70\,nm. 

The electrical (surface) structure of the device was fabricated on a 200\,$\mu$m-wide Hall bar. Contacts to both layers were established using AuGeNi Ohmic contacts. Electron-beam lithography was used to define a split gate (SG), a mid-line gate (MG), a barrier gate (BG) and a cut-off gate (CG)---used in setting up the tunnelling conditions---together with an array of wire gates (WG)---used in defining the experimental 1D system (see Fig.~\ref{fig:SF1_MM}A). The length of the wire gates was varied from 1--18\,$\mu$m. They were separated by 0.15--0.18\,$\mu$m gaps, and had a width of 0.1--0.3\,$\mu$m. A`parasitic' injection region also ran across the entire width of the mesa, with a fixed width of 0.45\,$\mu$m. A parasitic gate (PG) was used to modulate its density. All dimensions, particularly regarding the wire-region were carefully chosen in order to achieve minimal modulation of the lower-well carriers.

\subsection*{Momentum- and energy-resolved tunnelling spectroscopy}
\begin{figure*}
\centering
\includegraphics[width=0.99\textwidth]{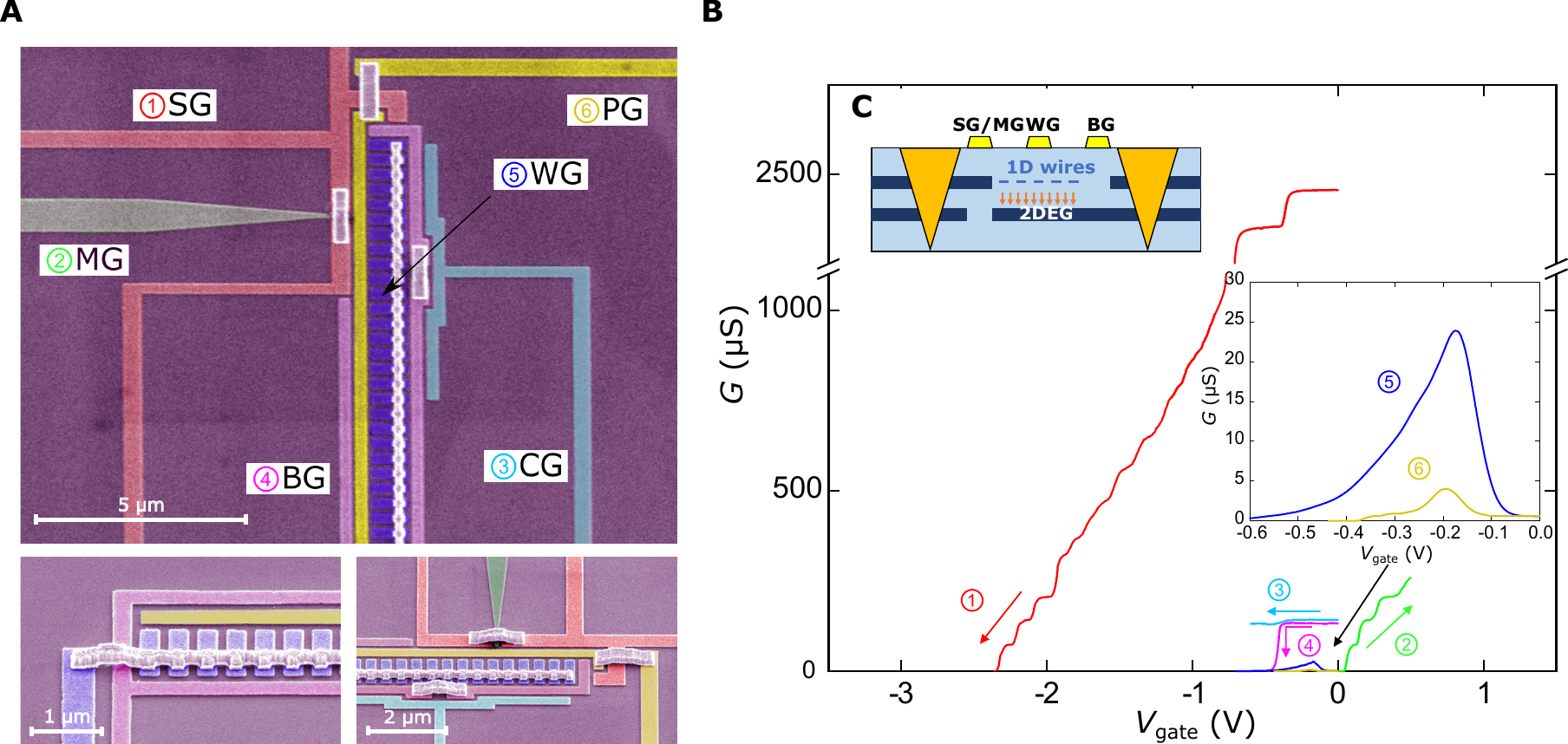}
\caption{\small \textbf{Vertical tunnelling device} (\textbf{A}) Scanning electron microscopy (SEM) images of the tunnelling device, showing the split (SG), mid-line (MG), barrier (BG), wire (WG) and parasitic (PG) gates. The cut-off (CG) gate is not used in this work and is always biased together with BG. Several samples were fabricated to vary the length of WG from 1--18\,$\mu$m (pictured, 1\,$\mu$m). The bottom micrographs show, respectively, the lower and upper ends of the wire array ($\sim$ 400 wires). In order to increase the uniformity of the 1D system, we developed a novel air-bridge technique to avoid having to use a connecting backbone structure (see \cite{jin_microscopic_2021}). (\textbf{B}) Gate operation and setting of tunnelling conditions. We start by negatively biasing SG (1), followed by positively biasing MG so that conductance is allowed only in the upper well (UW) (2). Next, we negatively bias both CG (3) and BG (4) but enough to only deplete the UW. Under this configuration, any signal measured between the ohmic contacts must result from direct tunnelling between each well. Inset: By varying WG and/or PG one can observe, respectively, 1D-2D and 2D-2D tunnelling between the wells (5 and 6). (\textbf{C}) Side profile of the tunnelling device. It consists of a double quantum-well heterostructure with a centre-to-centre distance of about $\sim$32\,nm.}
\label{fig:SF1_MM}
\end{figure*}

The tunnelling set-up was achieved as follows: first, the SG was negatively biased in order to pinch off both layers underneath, followed by positively biasing the MG in order to open a narrow conducting channel in the top well. At the other end of the device, the BG and the CG were biased enough to pinch off only the top layer. Under these conditions, any current injected through one of the Ohmic contacts had to have tunnelled between the layers in order to be detected (Fig.~\ref{fig:SF1_MM}B). 
 
Our spectroscopy technique consists of a low-noise, low-temperature measurement of the tunnelling current between the two 2D electron gas (2DEG) layers, which is given by \cite{Mahan_book}
\begin{align}
    I\propto\int {\textrm{d}\textbf{k}}\textrm{d}E\left[f_T(E-E_{\rm{F1D}}-eV_\textrm{DC})-f_T(E-E_{\textrm{F2D}})\right]\nonumber\\ 
    \times A_{1}({\bf k},E)A_{2}({\textbf{k}}+ed({\textbf{n}}\times{\textbf{B}})/\hbar,E-eV_{\textrm{DC}}),
\end{align}
where $e$ is the electron charge, $f_T(E)$ is the Fermi-Dirac distribution function, $d$ is the centre-to-centre well separation, {\bf n} is the unit normal to the 2D plane, ${\textbf{B}}=-B{\hat{\textbf{y}}}$ is the magnetic-field vector, ${\hat{\textbf{y}}}$ is the unit vector in the {\textit{y}}-direction, and $A_1({\textbf{k}},E)$ and $A_2({\textbf{k}},E)$ are the spectral functions of the 1D and 2D systems respectively, with the corresponding Fermi energies being  $E_{\textrm{F1D}}$ and $E_{\textrm{F2D}}$. The tunnelling current between the two layers is then proportional to the overlap integral of their spectral functions. We induce an offset $eV_{\textrm{DC}}$ between the Fermi energies of the two systems by applying a DC bias $V_{\textrm{DC}}$ between the layers. Similarly, an offset in momentum can also be obtained by applying a magnetic field of strength $B$ parallel to the 2DEG layers. When the field direction is along the (in-plane) $y$-direction, the Lorentz force then shifts the momentum of the tunnelling electrons in the $x$-direction by $edB$. Put together one can therefore map the dispersion of each system with respect to one another by measuring the differential conductance $G=\textrm{d}I/\textrm{d}V$ in both energy and momentum space.


\bibliography{references.bib}
\bibliographystyle{ScienceAdvances}

\noindent \textbf{Acknowledgements:} 
The authors would like to thank Leonid Glazman for assistance  and helpful comments. \\
\noindent \textbf{Funding:} This work was supported by the UK EPSRC [Grant Nos. EP/J01690X/1 and EP/J016888/1]. P.M.T.V. acknowledges financial support from EPSRC International Doctoral Scholars studentship via grant number EP/N509620/1, and the EPSRC Doctoral Prize. O.T. was funded by the DFG [project No. 461313466]. 
\\
\noindent \textbf{Author Contributions} P.M.T.V., Y.J., M.M., A.S.A, A.A. and W.K.T. fabricated the experimental devices, with P.M.T.V. and Y.J. performing the transport measurements shown. J.P.G. did the electron-beam lithography and I.F. and D.A.R. grew the heterostructure material.
P.M.T.V, O.T and C.J.B.F. analysed the data. O.T. and A.S. developed the theoretical framework. O.T. performed the calculations. C.J.B.F. supervised the experimental side of the project. All authors contributed to the discussion of the results. P.M.T.V., O.T. and C.J.B.F. wrote the manuscript.
\\
\noindent \textbf{Competing interests:} Authors declare that they have no competing interests.\\
\noindent \textbf{Data and materials availability:} All data needed to evaluate the conclusions in the paper are present in the paper and/or the Supplementary Materials. The data and modelling code that support this work are also available at the University of Cambridge data repository (DOI: https://doi.org/10.17863/CAM.81347).

\end{document}


\maketitle 

\textbf{This PDF file includes:}
\begin{itemize}
    \item Supplementary Text
    \item Fig. S1 to S9
\end{itemize}

\newpage

\section{Supplemental Text}

\subsection{Hubbard model} 
Interacting electrons in our quantum wires can be described by the 1D Fermi-Hubbard model, 
\begin{equation}
H=-t\sum_{j=1, \alpha=\uparrow,\downarrow}^{L/a}\left(c_{j\alpha}^{\dagger}c_{j+1,\alpha}+c_{j\alpha}^{\dagger}c_{j-1,\alpha}\right)+U\sum_{j=1}^{L/a}n_{j\uparrow}n_{j\downarrow},
\end{equation}
where $c_{j\alpha}$ are the second-quantisation operators in
real space obeying the Fermi commutation relations $\left\{ c_{j\alpha},c_{j'\alpha'}^{\dagger}\right\} =\delta_{jj'}\delta_{\alpha\alpha'}$, $\alpha$ is the spin-1/2 index $\uparrow$ or $\downarrow$, $n_{j\alpha}=c_{j\alpha}^{\dagger}c_{j\alpha}$
is the local density operator for each of the spin species, $L$
is the length of the wire, and $a$ is the lattice parameter of the host crystal. The two
microscopic constants of the model are the hopping amplitude $t$,
describing the kinetic energy, in the first term and the two-body interaction
energy $U$ in the second term.
The local nature of the density-density interaction in
this model allows for only a single microscopic constant $U$ that
takes into account at the same time the  strength of the
Coulomb interaction governed by the electronic charge $e$ and the screening
radius $R$, which we can tune in our experiment.

The many-body eigenstates of the Hubbard model in 1D were constructed
in \cite{Gaudin67,Yang67}. They describe
the amplitude of finding all $N$ particles at a given set of sites on the lattice $j_{1},\dots,j_{N}=\boldsymbol{j}$
and with a given configuration of their spins $\alpha_{1},\dots,\alpha_{N}=\boldsymbol{\alpha}$:
$\Psi=\sum_{\boldsymbol{j},\boldsymbol{\alpha}}a_{\boldsymbol{j}\boldsymbol{\alpha}}c_{j_{1}\alpha_{1}}^{\dagger}\cdots c_{j_{N}\alpha_{N}}^{\dagger}\left|0\right\rangle$. These
amplitudes have the form of a superposition of plane waves,
\begin{equation}
a_{\boldsymbol{j}\boldsymbol{\alpha}}=\sum_{P}A_{PQ\boldsymbol{\alpha}}e^{i\left( P\boldsymbol{k}a\right)\cdot \left(Q\boldsymbol{j}\right)},\label{eq:LiebWu_wave_function}
\end{equation}
where $Q$ is the permutation that orders all $N$ coordinates such that
\begin{equation}
Qj_{1}<\dots<Qj_{N},
\end{equation}
the momenta of $N$ particles are $\boldsymbol{k}=k_{1},\dots,k_{N}$,
and $\sum_{P}$ is the sum over all permutations of the charge momenta,
like in the Slater determinant for free particles. However, unlike
for free particles, the sign under the permutation of a pair of coordinates
is not $-1$ but is rather a phase factor $A_{PQ\boldsymbol{\alpha}}$ that also depends on
the spin configuration $Q\boldsymbol{\alpha}$ as 
\begin{multline}
A_{PQ\boldsymbol{\alpha}}=\left(-1\right)^{PQ}\sum_{R}\left(\prod_{1\leq l<m\leq M}\frac{R\lambda_{l}-R\lambda_{m}-\frac{iU}{2t}}{R\lambda_{l}-R\lambda_{m}}\right)\\ 
\times\prod_{l=1}^{M}\frac{\frac{iU}{2t}}{R\lambda_{l}-\sin Pk_{l}a+\frac{iU}{4t}}\prod_{j=1}^{s_{l}-1}\frac{R\lambda_{l}-\sin Pk_{j}a-\frac{iU}{4t}}{R\lambda_{l}-\sin Pk_{j}a+\frac{iU}{4t}},
\end{multline}
where $\boldsymbol{s}=s_{1},\dots,s_{M}$ are the coordinates of the $M$
spins $\uparrow$ in the configuration $\boldsymbol{\alpha}$ of all
spins of $N$ particles, $\boldsymbol{\lambda}=\lambda_{1},\dots,\lambda_{M}$ are the spin momenta that correspond to these $M$
spins $\uparrow$,
and $\sum_{R}$ is the sum over all permutations of these spin momenta. 

Spins of $N$ electrons in a quantum wire form a spin chain since the positions
of the electrons are ordered, permitting us to determine whether one
particular spin is to the left or right of another. For example, let
us consider a spin configuration for $N=8$ electrons, of which $5$
spins are $\downarrow$ and 3 spins are $\uparrow$, 
\begin{equation}
\alpha_{1}\dots\alpha_{N}=\downarrow\downarrow\uparrow\downarrow\uparrow\uparrow\downarrow\downarrow.
\end{equation}
Here the positions of spins $\uparrow$ are $\boldsymbol{s}=3,5,6$, which can be interpreted as coordinates of $M=3$ spin excitations in
the spin chain of length $N=8$. These spin positions can change when
electrons of opposite spin interact with each other via the Coulomb
interaction, changing their coordinates in the wire and making
the spin eigenmodes extended, at least at low densities away from the Mott-insulator regime. Therefore, $M$ spin excitations are  characterised by
$M$ spin momenta, like the charge particles. Moreover, electrons with the
same momenta can also interact, making their spins interact with each
other and causing the many-spin states consisting of $M$ individual spins  to be correlated in their own right.

The momenta are quantised by boundary conditions. Application of the periodic
boundary condition to the many-particle wave function in Eq.~(\ref{eq:LiebWu_wave_function}) gives
the Lieb-Wu equations \cite{LiebWu68},
\begin{align}
k_{j}L-\sum_{m=1}^{M}\varphi\left(\lambda_{m}-k_{j}a\right)&=2\pi I_{j},\label{eq:S_LiebWu_holons}\\
\sum_{j=1}^{N}\varphi\left(\lambda_{m}-k_{j}a\right)-\sum_{l=1}^{M}\varphi\left(\lambda_{m}/2-\lambda_{l}/2\right)&=2\pi J_{m},\label{eq:S_LiebWu_spinons}\\
\text{with}\quad\varphi\left(x\right)=-2\arctan\left(\frac{4tx}{U}\right),&
\end{align}
where $N$ non-equal integers $I_{j}$ and $M$ non-equal integers $J_{m}$ define the solution for the orbital $k_{j}$ and
the spin $\lambda_{m}$ momenta of an $N$-electron state for a given
value of the microscopic parameter $U/t$. This solution also gives
the eigenenergy of the many-electron state as  $E=ta^2\sum_{j=1}^{N}k_{j}^{2}$
and its momentum as $k=\sum_{j=1}^{N}k_{j}$. These simultaneous quantisation
conditions for both spin and charge degrees of freedom are a system
of $N+M$ connected nonlinear equations for any finite $U$. 

\begin{figure}[t]
\centering
\includegraphics[width=\textwidth]{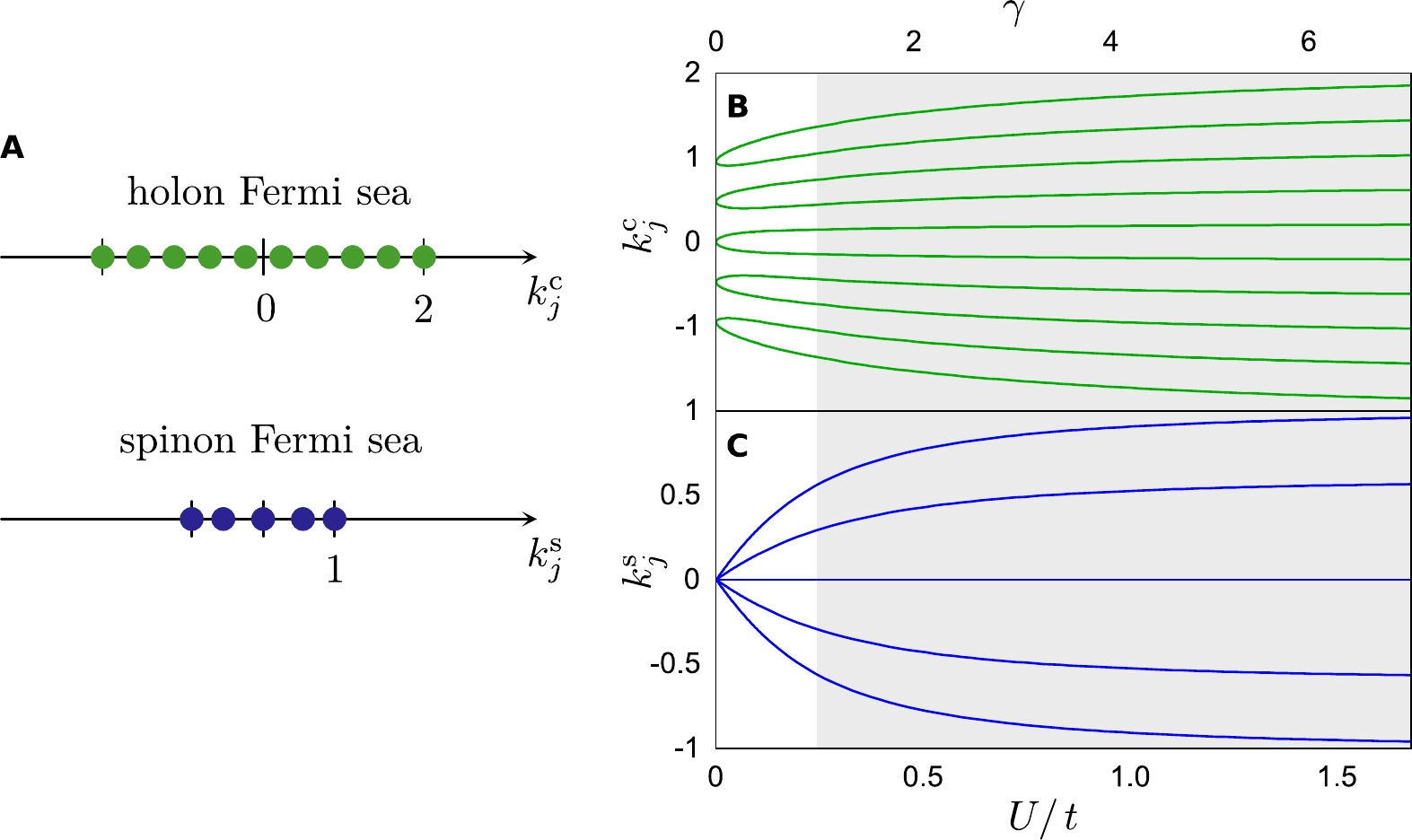}
\caption{\small \textbf{Ground state of the Fermi-Hubbard model.} (\textbf{A}) Solutions of the Lieb-Wu equations (\ref{eq:S_LiebWu_holons}) and (\ref{eq:S_LiebWu_spinons}) for the charge and the spin degrees of freedom for the ground state in the infinite interactions limit $U=\infty$ and its evolution for different values of $U$ for the holon (\textbf{B}) and the spinon (\textbf{C}) Fermi seas for $N=10$ unpolarised $M=5$ electrons. 
The charge momenta are normalised by the free-electron Fermi momentum,  $k^\textrm{c}_j=k_j/k_\textrm{F}$ where $k_\textrm{F}=\pi N/\left(2L\right)$ and the spin momenta  $k^\textrm{s}_j=-2\arctan[U/(4t\lambda_j)]/\Lambda$ are normalised by the solution of the integral Orbach equation $\Lambda$ \cite{Orbach58} that plays the role of the Fermi momentum for Heisenberg spin chains. The dimensionless parameter $\gamma$ is defined in Eq.~(\ref{eq:S_gamma_discrete}). In the grey area, $\gamma>1$, the interactions are strong enough for the two Fermi seas to be fully developed.}
\label{fig:SF_two_seas}
\end{figure}


\subsection{Two Fermi seas}

Both the charge and the spin momenta correspond to non-equal integer numbers $I_{j}$ and $J_{m}$ since the 1D wave function in Eq.~(\ref{eq:LiebWu_wave_function}) becomes zero for any pair of equal $k_j$ or $\lambda_m$, which are obtained immediately from Eqs.~(\ref{eq:S_LiebWu_holons}, \ref{eq:S_LiebWu_spinons}) when a pair of  $I_{j}$ or $J_{m}$ are equal. In principle, such an emergent
picture is described by two different Fermi seas, in which both kinds of modes are filled with spin or charge excitations up to some finite
densities, just as the usual Fermi sea is filled with non-interacting fermions up to a finite Fermi energy.
The difference for the problem with interactions is a more complicated relation
between the two types of the occupation numbers, visualised by a set of $I_{j}$ and $J_{m}$,
and the actual distributions of the two types of momenta, $k_{j}$
and $\lambda_{m}$. This relation is not just a linear quantisation condition for each momentum of the free particles independently but is rather
given by the solution of the system of essentially nonlinear coupled equations (\ref{eq:S_LiebWu_holons}) and (\ref{eq:S_LiebWu_spinons}) for any finite interaction $U$. For instance, the quantisation conditions given by the Lieb-Wu equation in Eqs.~(\ref{eq:S_LiebWu_holons}, \ref{eq:S_LiebWu_spinons})
alter qualitatively the equilibrium properties by changing 
parameters of the two Fermi seas or by changing the excitation energies
of different modes, making their dispersions generally interaction-dependent. 

Our quantum wires remain unpolarised up to the maximum field of $7\,$T
that we use in our experiments.  The number of spin-up and
down electrons is equal within 10\% (5\% below $B_{+}$), and we therefore assume $M=N/2$ spin particles.
For the holon Fermi sea, which has twice as many filled states as the spinon
Fermi sea, the Lieb-Wu equations give the two distributions of their
momenta at the infinite-interaction point $U=\infty$, see Fig.~\ref{fig:SF_two_seas}A. We define the Fermi
momentum $k_{\textrm{F}}$ by the free electrons, \emph{i.e.}\ 
by $N/2$ fermions that are doubly degenerate with respect to spin
without the interactions at $U=0$. For strong enough interactions the free-electron
degeneracy is completely lifted, increasing the density of the charge
particles by the factor of two. At the same time, the spin particles
develop the second Fermi sea of the Heisenberg-spin-chain type \cite{Ogata90} with a density that is twice as small as that of the holons. The momentum distribution in the spinon Fermi sea is non-equidistant: the spin momenta have a pronounced higher density towards the Fermi points but are sparser at the bottom of the band \cite{Gaudin_book}. A Fermi sea of this kind was observed using neutron scattering in non-itinerant magnets \cite{Lake05,Mourigal13,Lake13}.

Decreasing the interaction energy $U$ away from the infinite point, we solve the Lieb-Wu
equations (\ref{eq:S_LiebWu_holons}) and (\ref{eq:S_LiebWu_spinons}) numerically, see Figs.~\ref{fig:SF_two_seas}B and C. Down to an intermediate value of $U$ both Fermi seas remain stable, but below it the two-seas picture starts to change qualitatively.
For weak interactions the holon Fermi sea becomes doubly degenerate,
recovering the free-fermion picture, and the spinon Fermi sea collapses,
becoming the spin part of the free-electron function that describes
$\pm1$ permutation signs for the electrons of different or the
same spin. 

The dimensionless parameter $\gamma$, which controls this transition, emerges
from the Hubbard model itself microscopically. It was identified in
a quantitative analysis of the double occupancy of the electronic states \cite{OT14}
as 
\begin{equation}
\gamma=\frac{\lambda_\textrm{F}}{16a}\frac{U}{t}\frac{1}{1-\frac{1}{N}\sum_{l=1}^{N/2}\frac{\lambda_{l}^{2}\left(\infty\right)-\left(\frac{U}{4t}\right)^{2}}{\lambda_{l}^{2}\left(\infty\right)+\left(\frac{U}{4t}\right)^{2}}},\label{eq:S_gamma_discrete}
\end{equation}
where $\lambda_\textrm{F}=4L/N$ is the Fermi wavelength of the free-electron
gas, $a$ is the lattice parameter, and $\lambda_{l}\left(\infty\right)$
are the spin part of the solution of Eqs.\ (\ref{eq:S_LiebWu_holons}) and (\ref{eq:S_LiebWu_spinons}) in the infinite-interaction
limit $U=\infty$. Turning the sum into an integral in the thermodynamic limit
and evaluating it for the unpolarised Heisenberg chain \cite{Orbach58}
we obtain the numerical value of the denominator in the above expression,
$1-\sum_{l}\dots/N=1.193(1)$. Thus, in the thermodynamic limit the
parameter controlling the interaction effects is
\begin{equation}
\gamma=0.032\frac{\lambda_\textrm{F}}{a}\frac{U}{t}.\label{eq:S_gamma_thremodynamic_limit}
\end{equation}
When the interactions are strong, $\gamma>1$, both the spinon and
the holon Fermi seas are fully developed but, when $\gamma<1$, the
holon Fermi sea is still close to the double occupancy of the free Fermi
gas. The expression in Eq.\ (\ref{eq:S_gamma_thremodynamic_limit})  is presented in Eq.\ (1) in the main text.

\subsection{Spin and charge excitations}

\begin{figure}[t]
\centering
\includegraphics[width=0.67\textwidth]{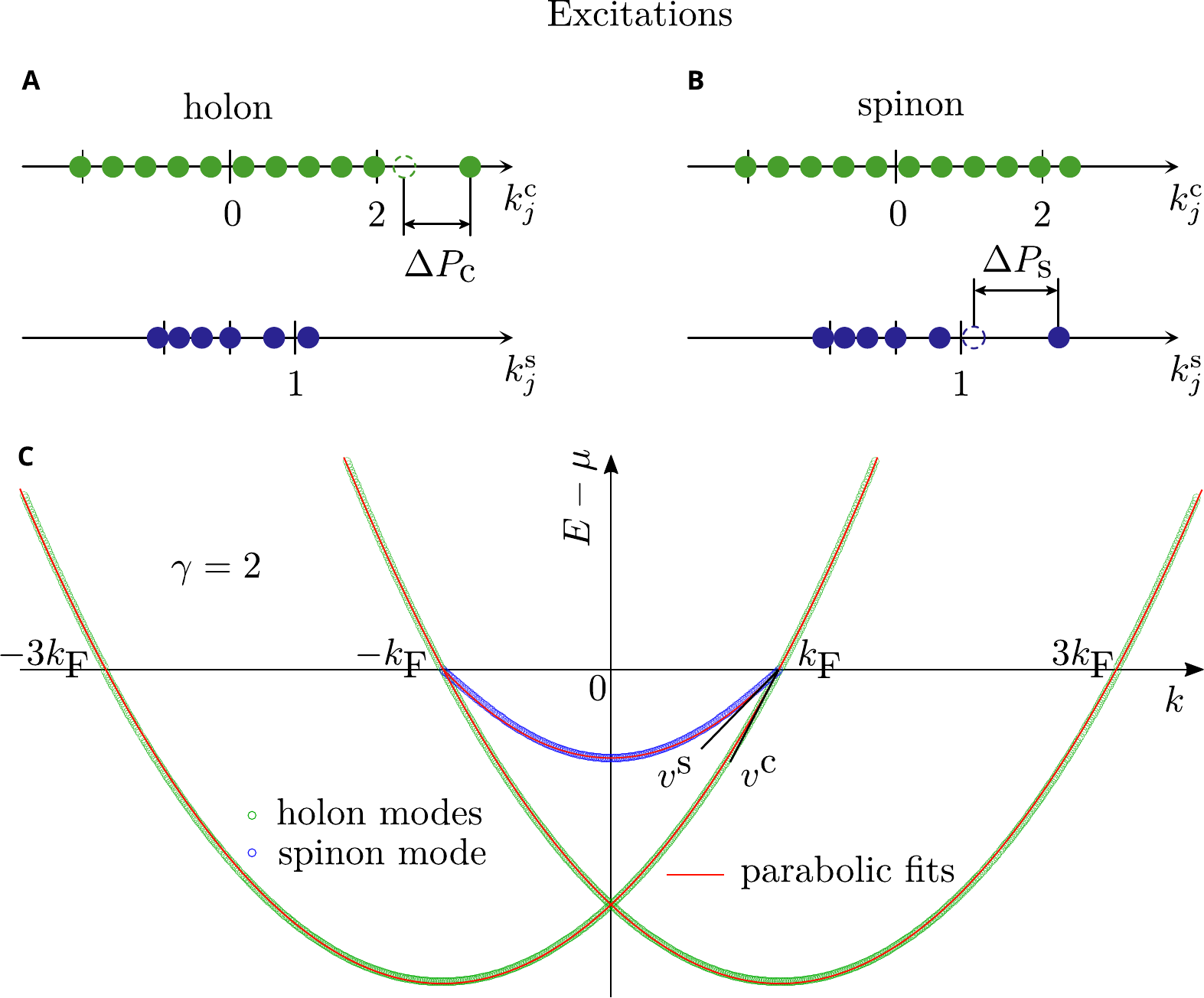}
\caption{\small \textbf{Spinon and holon excitations of Fermi-Hubbard model.} (\textbf{A}) Excitation of the holon type, in which the spinon part is placed at the lowest possible momentum for the spin excitations and the holon part is promoted to some finite momentum, which is larger than the lowest possible momentum for the charge excitations by $\Delta P_\textrm{c}$. (\textbf{B}) Excitation of the spinon type, in which the holon part is placed at the lowest possible momentum for the charge excitations and the spin part is promoted to some finite momentum, larger than the lowest possible momentum for the spin excitations by $\Delta P_\textrm{s}$. (\textbf{C}) The dispersions of the holon (green circles) and of the spinon (blue circles) modes consisting of the excitations in (\textbf{A}) and (\textbf{B}), respectively, obtained by solving the Lieb-Wu equations (\ref{eq:S_LiebWu_holons}, \ref{eq:S_LiebWu_spinons}) for $N=550$ unpolarised electrons with $M=275$ for interaction strength $\gamma=2$. The red lines are the parabolic fits for both modes. The two black lines at the Fermi point $k_\textrm{F}$ mark the linear dispersions of holons $v^\textrm{c}$ and of spinons $v^\textrm{s}$ of the linear spinful Tomonaga-Luttinger model.}
\label{fig:SF_hs_excitations}
\end{figure}

By adding a single electron with its charge and spin we add two excitations
to the system simultaneously, one holon and one spinon. With  two
Fermi seas there are two options. One is adding a holon to its Fermi sea at the
$2k_{\textrm{F}}$ point and spinon to its Fermi sea at $-k_{\textrm{F}}$
point on top of the ground state in Fig.\ \ref{fig:SF_two_seas}A, obtaining the net momentum for the electronic excitation $k=k_{\textrm{F}}$.
Another possibility is adding the holon and the spinon at the same
sides of their Fermi seas, producing the net momentum $k=3k_{\textrm{F}}$.
Our wires have inversion symmetry. Therefore, the mirror of this
argument produces the $k=-k_{\textrm{F}}$ and the $k=-3k_{\textrm{F}}$
as well, see Fig.\ \ref{fig:SF_hs_excitations}C. 

Starting at a Fermi point, say $k_\textrm{F}$, we can add a pair of excitations
of the two kinds not just at the Fermi points of their respective Fermi seas but also at some finite momenta,
$\Delta P_\textrm{c}$ or $\Delta P_\textrm{s}$, above them, see Fig.\ \ref{fig:SF_hs_excitations}A
and B. Change of the momentum of any of these excitations distorts
both Fermi seas via solving the nonlinear
Lieb-Wu equations (\ref{eq:S_LiebWu_holons}) and (\ref{eq:S_LiebWu_spinons}): all states in the spinon Fermi sea become
generally asymmetric, in addition to being non-equidistant in the
ground state, and the holon Fermi sea as a whole shifts by the total spinon
momentum, see Figs.\ \ref{fig:SF_hs_excitations}A and B. However, the total momentum of the 
electronic excitation is a quantum number, which has a simple relation,
$k=2\pi\left(\sum_{j}I_{j}+\sum_{m}J_{m}\right)/L$, to the whole
set of the quantisation integer numbers, $I_{j}$ and $J_{m}$, and
is independent of $U$ \cite{LiebWu68}. This result of the total momentum
conservation law for a many-body system defines the momentum of the
holon and the spinon excitations as 
\begin{equation}
k=k_{\textrm{F}}+\Delta P_\textrm{c}\quad\textrm{and}\quad k=k_{\textrm{F}}+\Delta P_\textrm{s},
\end{equation}
where $\Delta P_\textrm{c}=2\pi I_{N+1}/L$ and $\Delta P_\textrm{s}=2\pi J_{M+1}/L$
can be visualised in a simple way due to the same total momentum conservation for all particles. Therefore, the distance in the momentum variable between the neighbouring points is always equal to the quantum of  momentum $2\pi/L$ and the position of the second crossing points of the holon modes with the line of chemical potential in Fig.\ \ref{fig:SF_hs_excitations}C always remains at the same momentum $\pm 3k_{\textrm{F}}$, irrespective of the value of $U$.

The energies of these excitations, on the other hand, are more complicated.
In order to evaluate them and the dispersions that they form, we need
to solve the Lieb-Wu equations (\ref{eq:S_LiebWu_holons}) and (\ref{eq:S_LiebWu_spinons}) numerically for different
values of $U$, see Fig.\ \ref{fig:SF_hs_excitations}C. At low energies, the solutions form two linear
dispersions with non-commensurate slopes, $v^{\textrm{c}}$ and $v^{\textrm{s}};$
the stronger $U$ is, the larger is the ratio $v^{\textrm{c}}/v^{\textrm{s}}$,
and the two velocities become equal $v^{\textrm{c}}/v^{\textrm{s}}=1$ in the free-particle limit $U=0$. These two holon
and spinon velocities are nothing but a pair of the Luttinger parameters
in the spinful linear Tomonaga-Luttinger model \cite{Coll74,Schultz90} at low energies. They are manifested in the observable
spectral function as a pair of divergences \cite{Schoenhammer92,Voit93}.

Extensions of the momenta of the excitations of two different kinds, $\Delta P_\textrm{c}$ and
$\Delta P_\textrm{s}$, away from the $\pm k_{\textrm{F}}$ points provide a natural
generalisation of these linear modes to the nonlinear regime, see Fig.\ \ref{fig:SF_hs_excitations}C. Unlike
the linear regime, where there are only two branches around each of the
Fermi points, in the whole energy band there are three. This is a
manifestation of the different densities of the holon and the spinon
Fermi seas. Dispersions of all of the three curves are not exactly parabolic for
finite $U$ but are close to parabolae due to an interaction effect between the two Fermi seas, with the most significant deviations occurring for the spinon branch in the vicinity of the Fermi points. Fitting parabolae to the numerically
produced exact dispersions (red lines in Fig.\ \ref{fig:SF_hs_excitations}C), gives the dependence of the two
masses $m_{\textrm{c}}$ and $m_\textrm{s}$  on the interaction parameter $\gamma$. Note that for a finite $\gamma>0$ these two masses are non-commensurate but they become commensurate, $m_c=2m_s$, in the $\gamma=0$ limit. The ratio of these masses is presented in Fig. 4B\ of the main text. 

The two velocities $v^{\rm c}$ and $v^{\rm s}$ are extracted as the linear coefficients in the dispersion of the holon and the spinon modes at the Fermi energy, and the dependence of their ratio on $\gamma$ is also presented in Fig.\ 4B. We fit our experimental data with the exact dispersions of the Hubbard model to obtain the interaction parameter $\gamma$ directly, and also with the two parabolae, see more details in the section `Parabolic Model' below.

\subsection{Mode hierarchy away from the Fermi points}

\begin{figure}[t]
\centering
\includegraphics[width=0.55\textwidth]{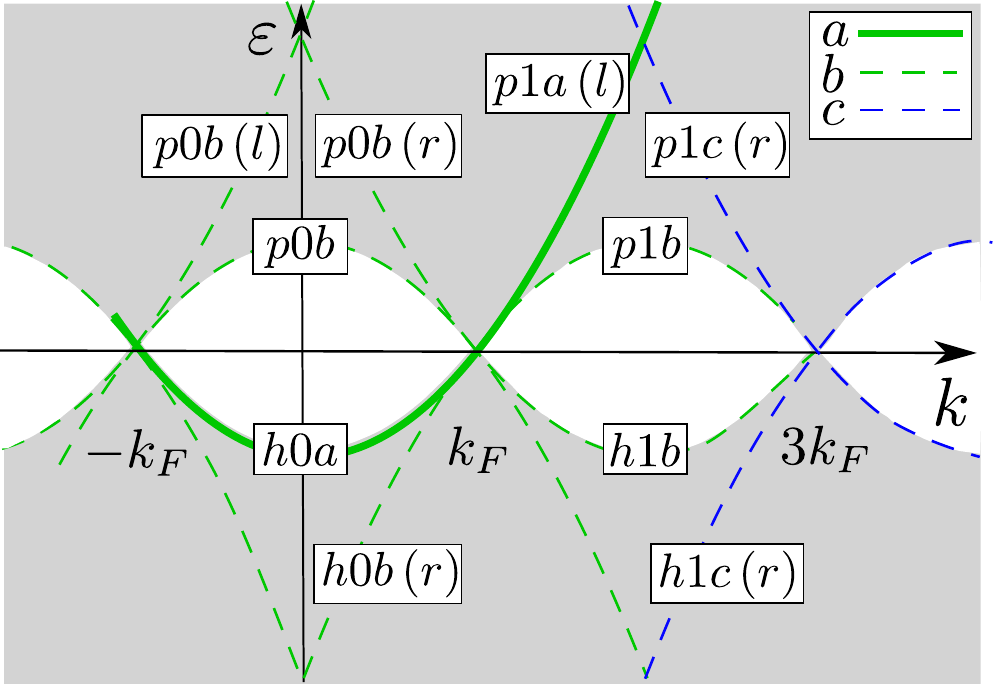}
\caption{\small \textbf{Spectral function of spinless fermions according to the mode-hierarchy picture.} First (second)-level modes are shown in the region $-k_{\textrm{F}}<k<k_{\textrm{F}}$ ($k_{\textrm{F}}<k<3k_{\textrm{F}}$) and are labelled by 0 (1), where $k_{\textrm{F}}$ is the Fermi wave vector. Accessible and forbidden regions are marked by grey and white, respectively. Particle (hole) sectors are marked by $p$ ($h$) for positive (negative) energies, $a$, $b$ and $c$ correspond respectively to the level in the mode hierarchy in powers of 0, 1 and 2 of $R^2/L^2$ and $(r,l)$ correspond to the origin in the range, right or left.}
\label{fig:hierarchy_spinless}
\end{figure}

This section follows the work in \cite{Tsyplyatyev15,Tsyplyatyev16} previously described in \cite{vianez_chapter_2021}. A systematic understanding of the general picture for nonlinear excitations
has come from a microscopic analysis of the spinless counterpart of the
1D Fermi-Hubbard model. Using the analysis of the correlation functions
via the algebraic Bethe ansatz, it was found \cite{Tsyplyatyev15,Tsyplyatyev16}
that away from the Fermi points the exponentially many excitations forming the many-body continuum are separated into
levels of a mode hierarchy according to their spectral strength, which
is proportional to integer powers of a small parameter $R^{2}/L^{2}$,
where $R$ is the radius of the two-body interaction potential and
$L$ is the length of the system. A detailed analysis of these amplitudes
revealed that the
strongest excitations (with the zeroth power of $R^{2}/L^{2}$) form a parabola-like dispersion, which corresponds
to the solid green line in Fig.\ \ref{fig:hierarchy_spinless}, similar to that of the original non-interacting
fermions but with some renormalisation due to interactions. All
other many-body modes have powers of $R^{2}/L^{2}$ greater than zero,
with the general trend of increasing the integer by one with each
discrete step of $2k_{\textrm{F}}$ along the $k$-axis away from the principal parabola, see the dashed lines in Fig.\ \ref{fig:hierarchy_spinless} and their labelling scheme in the caption to the figure.

As an example, the hole part of the principal parabola, between the
$\pm k_{\textrm{F}}$ points, has the largest amplitude but its mirror
in the particle sector---a `replica' in the shape of a dome marked as $p0b$ by the dashed green in Fig.\ \ref{fig:hierarchy_spinless}---has a parametrically smaller amplitude, proportional
to the first power of $R^{2}/L^{2}$. In the spectral function the
strength of this replica is predicted to be 
\begin{equation}
A_{1}\left(k,E\right)\propto\frac{R^{2}}{L^{2}}\frac{k_{\textrm{F}}^{2}k^{2}}{\left(k^{2}-k_{\textrm{F}}^{2}\right)^{2}}\delta\left(E-\mu+\xi_{1}\left(k\right)\right),
\end{equation}
where $\xi_{1}\left(k\right)$ is the dispersion for the `replica'.
Therefore, for almost all momenta this mode will be unobservable in
the thermodynamic limit. The only exceptions are the regions around
the $\pm k_{\textrm{F}}$ points, where the singularity in the denominator
starts to compete with the parametric smallness, resulting in a large
amplitude overall. On the other hand, the measurement of the whole
mode requires using smaller systems, in which the $R^{2}/L^{2}$ parameter
still leaves the amplitude of the whole mode above the background
from the other processes.

Close to the spectral threshold the mode hierarchy reproduces the
predictions of the mobile-impurity model. The small parameter $R^{2}/L^{2}$
provides a path for accounting for the principal amplitudes of the
exponential continuum of many-body excitations, which leads to the
microscopic calculation of the spectral function. The resulting threshold
exponents match those predicted by the phenomenologically introduced
nonlinear hydrodynamic model.

While the mode hierarchy emerges away from the Fermi points, close
to them (where the spectrum is almost linear) it transitions into
the usual linear TLL, see details in \cite{Tsyplyatyev16}. The hydrodynamic
modes of the latter consist of a huge number of many-body modes, all
of which are of similar spectral strength, making the two regimes
distinct already on the microscopic level. The change from one into
the other can be traced quite easily using a macroscopic quantity,
the density of states. It can be calculated exactly by numerical means
using the Bethe ansatz approach, exhibiting the power-law suppression around the
Fermi energy $E_{\textrm{F}}$ predicted by the TLL model and a crossover
into a finite density $\propto1/\sqrt{E}$ predicted by the mode hierarchy
away from the linear region, where the nonlinearity of the single-particle
dispersion already destroys the hydrodynamic modes of the TLL.

In our experiment, each mode of the hierarchy should, in principle, be split into two by the effect of spin-charge separation. We start observing the spin part of this p0b `replica' mode in wires of 5$\mu$m and shorter lengths, see region C in Fig.\ 3A and Fig.\ 3C in the main text, whereas the p1b replica, see region B in Fig.\ 3A, is visible at 5$\mu$m, and strengthens as the length of the wire decreases, as shown in  Fig.\ 3B in the main text.

\subsection{Interaction parameter $r_\textrm{s}$ in 1D}

The Wigner-Seitz radius $r_{\textrm{s}}$ is used as a generic interaction
parameter in Fermi systems independent of their dimensionality. Its common
applicability stems from the fact that the Fermi energy is inversely
proportional to the inter-particle distance squared and that the Coulomb energy is inversely proportional to the same inter-particle
distance but to the first power. Therefore, the dimensionless ratio
of the interaction energy to the kinetic energy is just given by $r_{\textrm{s}}$. 

\begin{figure*}[ht!]
\centering
\includegraphics[width=0.9\textwidth]{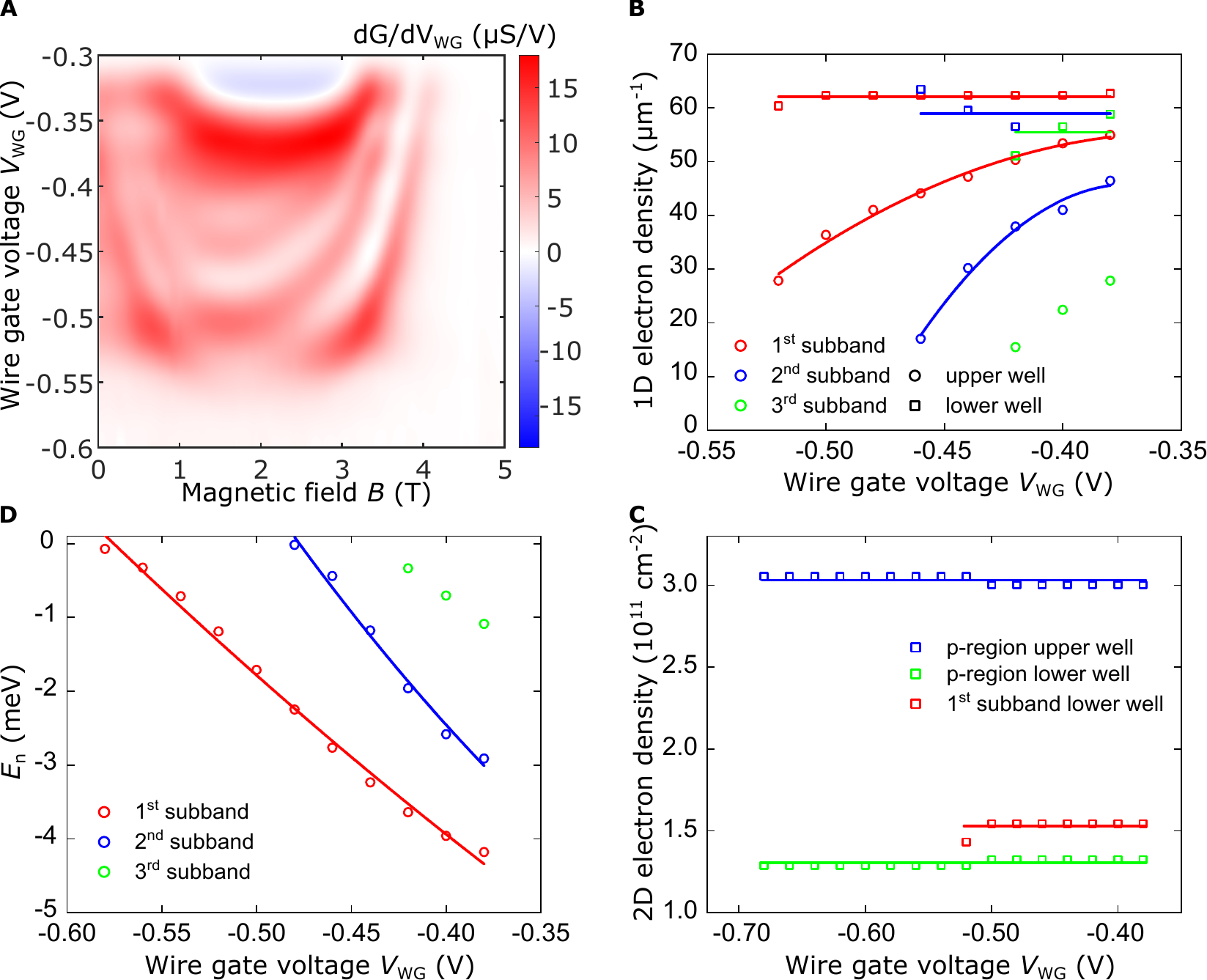}
\caption{\small \textbf{1D wire subbands.} (\textbf{A}) $\textrm{d}G/\textrm{d}V_{\textrm{WG}}$ differential of the tunnelling conductance $G$ with respect to the wire-gate voltage $V_{\textrm{WG}}$, as a function of $V_{\textrm{WG}}$ and magnetic field $B$ perpendicular to the wires, as obtained under equilibrium conditions $V_{\textrm{DC}}=0$, for a 1.7\,$\mu$m device. Three fully developed 1D subbands can be seen below the 2D band at $V_{\textrm{WG}}\approx -0.35$\,V and above the cutoff point at $V_{\textrm{WG}}\approx -0.55$\,V, below which the wires cannot conduct. (\textbf{B}) Equilibrium 1D electron densities in both upper (circles) and lower (squares) wells for each occupied subband. Note that the electrons are not laterally confined in the bottom well and as such the lower well 1D density has no physical meaning and is only shown for comparison. (\textbf{C}) 2D electron densities in both upper (blue) and lower (green) wells' `parasitic' regions. The overall independence of the 2D densities on the wire-gate voltage and the proximity of the (now physical) 2D electron density of the bottom well below the wire region (red) to that of the injection region shows that the bottom 2DEG remains largely unaffected by $V_{\textrm{WG}}$, and as such can be taken to be a well-understood 2D probe. (\textbf{D}) Subband energies relative to the chemical potential $\mu=0$ as a function of $V_{\textrm{WG}}$.}
\label{fig:SF_FG}
\end{figure*}

We can extract the value of the Fermi wavelength $\lambda_\textrm{F}$
in our experiment from the density of the 1D system, which can be obtained from either the 1D wire subbands dispersion (\emph{e.g.} see Fig.\ \ref{fig:SF_FG}) or alternatively by using the zero-field intersect points in the tunnelling conduction maps (see section on `Parabolic Model'). This gives us two independent estimates for the electron density. Its relation to $r_{\textrm{s}}$ in 1D is as follows.
In one dimension the electron density is expressed in terms of $r_{\textrm{s}}$
as 
\begin{equation}
n=\frac{1}{2a_{\rm B}r_{\textrm{s}}},\label{eq:S_n_rs}
\end{equation}
where 
\begin{equation}
a_{\rm B}=\frac{4\pi\varepsilon\varepsilon_{0}\hbar^{2}}{me^{2}}
\end{equation}
is the Bohr radius of electrons in GaAs, with $\varepsilon\approx 12$ and $m=0.067m_{\rm e}$. On the other hand, the density of free electrons in
a 1D Fermi sea is also given by the integral over the Fermi function,
\begin{equation}
n=2\int\frac{dk}{2\pi}n_{k}=\frac{4}{\lambda_{\rm F}},\label{eq:S_n_lambda_F}
\end{equation}
where the occupation numbers are a step function, $n_{k}=\theta\left(k_{\textrm{\rm F}}-\left|k\right|\right)$,
and $k_{\textrm{F}}=2\pi/\lambda_{\textrm{\rm F}}$. Combining Eqs.\ (\ref{eq:S_n_rs}) and (\ref{eq:S_n_lambda_F}) produces the following relation 
\begin{equation}
r_{\textrm{s}}=\frac{\lambda_{\rm F}}{8a_\textrm{B}}.
\end{equation}
This generic dimensionless interaction parameter $r_\textrm{s}$ plays the same role as the
microscopic dimensionless interaction parameter of the 1D Fermi-Hubbard $\gamma$ in Eq.~(\ref{eq:S_gamma_discrete}). However, $r_\textrm{s}$ does not account for screening effects, see the main text.

The 1D subband structure of electrostatically-gated wires such as the ones used in this experiment can be observed from Fig.\ \ref{fig:SF_FG}A. As can be seen, this is relatively simple, with our measurements being in very good agreement with the predictions and simulations of \cite{laux_quasi_one_dimensional_1988} and \cite{owen_ground_state_2016}. In specific, \cite{owen_ground_state_2016} took into account the explicit effect of e-e interactions into the (sub)band structure to some extent, finding no significant change in the overall structure. This, together with an observed subband spacing of $\sim2$\,meV (see also Fig.\ \ref{fig:SF_FG}D), allows us to individually populate each subband which should, therefore, not strongly hybridise. This simple (sub)band structure is in full agreement with what is observed in the tunnelling maps, both when in the 2D-2D and 1D-2D regimes, see section 1.6. for more details.

\subsection{Parabolic model}

When fitting the 1D tunnelling resonances observed in the dispersion maps, one can approximate the exact numerical solutions arising from the Hubbard model by parabolae, see Supplementary Text `Spin and charge excitations' and Fig.\ \ref{fig:SF_hs_excitations}C for both spin- or charge-type excitations. On the other hand, 2D systems at low densities (see Fig.\ \ref{fig:SF_FG}C), are known to be Fermi liquids with effective mass renormalised by interactions. Together these account for the dispersion of the elementary excitations in both the upper and lower wells (UW and LW), which can behave as either 2D or 1D systems depending on whether the signal arises from unconfined (`parasitic'-, LW) or confined (wire-) regions of the device. We will now show how separation between spin- and charge-type modes in 1D, each associated with different effective masses and degeneracies, emerges naturally from the data after considering simpler models. We start by introducing the capacitive-coupling correction, which was also applied when fitting using the Hubbard model.

\subsubsection{Modelling the capacitive coupling}

Tunnelling between the upper and lower layers of a device, across its dielectric barrier, is affected by capacitance effects, which result in the observed dispersions being slightly asymmetric. Owing to the finite capacitance $C$ between the two layers, a small increase or reduction of the electron density $\pm\delta n_{\textrm{2D/1D}}$ occurs at each side of the barrier. We have $e\delta n_{\textrm{2D}}=V_{\textrm{DC}}C/A$ for the 2D system (of area $A$) and $e\delta n_{\textrm{1D}}=V_{\textrm{DC}}C/L$ for the corresponding 1D system (of total length $L$), with $V_{\textrm{DC}}$ the DC-bias applied between the wells. For two layers with Fermi wavevectors $k_{\textrm{F,1}}$ and $k_{\textrm{F,2}}$, where $k_{\textrm{F,1}}<k_{\textrm{F,2}}$, the zero-bias crossing points $B_+$ and $B_-$ can be combined to give
\begin{equation}\label{layer_fermi_vectors}
    k_{\textrm{F,1}}=\frac{ed}{2\hbar}(B_+-B_-), \hspace{1cm} k_{\textrm{F,2}}=\frac{ed}{2\hbar}(B_++B_-),
\end{equation}
where $d$ is the separation between the wells. Our 1D wires, once defined by the gate voltage, do indeed have lower density than the 2D system beneath, so in the wire region we take $k_{\textrm{F,1}}$ as the Fermi wavevector of the 1D system in the absence of interactions, and $k_{\textrm{F,2}}$ as that of the 2D layer.

The electron densities for 1D and 2D systems are given by
\begin{equation}\label{1D_density}
    n_{\textrm{1D}}=\frac{\textrm{d}N}{\textrm{d}L}=\frac{2k_{\textrm{F,1D}}}{\pi}
\end{equation}
and
\begin{equation}\label{2D_density}
    n_{\textrm{2D}}=\frac{\textrm{d}N}{\textrm{d}A}=\frac{k_{\textrm{F,2D}}^2}{2\pi}.
\end{equation}

From here, we get
\begin{equation}
    \delta n_{\rm 1D}=n_{\rm 1D}(V_{\textrm{DC}})-n_{\rm 1D}(0)=\frac{\eta_i C V_{\textrm{DC}}}{eL}=\frac{2}{\pi}(k_{\textrm{F,1D}}(V_{\textrm{DC}})-k_{\textrm{F,1D}}(0))
\end{equation}
and
\begin{equation}
    \delta n_{\rm 2D}=n_{\rm 2D}(V_{\textrm{DC}})-n_{\rm 2D}(0)=\frac{\eta_i C V_{\textrm{DC}}}{eA}=\frac{1}{2\pi}(k_{\textrm{F,2D}}^{2}(V_{\textrm{DC}})-k_{\textrm{F,2D}}^{2}(0)),
\end{equation}
which in turn gives the modified Fermi wavevectors as the inter-layer voltage $V_{\textrm{DC}}$ causes the densities to change,
\begin{equation}\label{S6}
    k_{\textrm{F,1D}}(V_{\textrm{DC}})=k_{\textrm{F,1D}}(0)+\frac{\pi\eta_{i}C V_{\textrm{DC}}}{2eL}
\end{equation}
and 
\begin{equation}\label{S7}
    k_{\textrm{F,2D}}(V_{\textrm{DC}})=\sqrt{k_{\textrm{F,2D}}^{2}(0)+\frac{2\pi\eta_iC V_{\textrm{DC}}}{eA}}.
\end{equation}
Here $\eta_{i}=\pm1$, with $i={1,2}$ labelling the upper/lower layer respectively, is a sign factor based on the experimental setup. In our experiment, $\eta_{\textrm{1}}=-1$ and $\eta_{\textrm{2}}=1$, since for $V_{\textrm{DC}}>0$ the upper well was more positive than the lower well.

The bottom well is always 2D in nature, which means that  $k_{\textrm{F,2}}=k_{\textrm{F,2D}}(V_{\textrm{DC}})$ with $\eta_2=1$. The top well, on the other hand, can behave as being either 2D, in the injection region, or 1D, in the wires. For the latter we have $k_{\textrm{F,1}}=k_{\textrm{F,1D}}(V_{\textrm{DC}})$ with $\eta_1=-1$. For the former, the upper-well density is higher than in the lower well.  Note also that equation \eqref{S7} reduces to equation \eqref{S6} when expanded in the low-capacitance limit,
\begin{align}
    k_{\textrm{F,2D}}^{'}&=k_{\textrm{F,2D}}\bigg(1+\frac{2\pi\eta_i C V_{\textrm{DC}}}{eAk_{\textrm{F,2D}}^2}\bigg)^{1/2}\\ 
    &\approx k_{\textrm{F,2D}}+\frac{\pi\eta_i C V_{\textrm{DC}}}{eAk_{\textrm{F,2D}}} .\label{S9}
\end{align}
Therefore, comparing Eqs.\ \ref{S6} and \ref{S9} we conclude that the capacitance correction for the wire region, which is 1D in nature, can nevertheless be treated as for a 2D system by transforming the capacitance as $C/L=C/A\times 2/k_{\textrm{F,1D}}$. One can also assume charge neutrality of the pair of layers to estimate the width of the wires $w$ by solving $C/L=w\times C/A$.

We now estimate the capacitive coupling between the two wells in our setup, when applying a voltage to the upper layer and keeping the potentials of the gates and lower layer fixed. This is required to correct for the mapped dispersions as discussed previously. In a classical Coulomb system, the capacitance $C\equiv \textrm{d}Q/\textrm{d}V_{\textrm{DC}}$ of a conductor is a purely geometric quantity (for charge $Q$ on the conductor). For example, in a parallel-plate capacitor, it takes the well-known value $C=\varepsilon\varepsilon_0 A/d$, $A$ being the surface area of the plates, $d$ their separation, and $\varepsilon$ the relative permittivity of the dielectric material in between. In systems with a low density of states however, such as a 2DEG, this result does not strictly apply, since unlike for a perfect metal plate, the density of states here is no longer infinite. Therefore, in contrast to a metal, a 2DEG cannot, in general, perfectly screen the electrical field generated by the surface gates.

In order to account for this effect one has to consider band-filling/band-emptying in changing the density of states in the 2DEG as the gate voltage is varied. This correction was initially proposed in \cite{luryi_quantum_1988} and can be modelled by considering two capacitors connected in series,
\begin{equation}
    \frac{1}{C}=\frac{1}{C_{\rm G}}+\frac{1}{C_{\rm Q}},
    \label{capacitances}
\end{equation}
where $C_{\rm G}$ is the usual geometric capacitance, $C_{\rm Q}=e^2dn_{\rm 2D}/dE_{\rm F}A$ is the new quantum capacitance, 
and $E_{\rm F}$ is the Fermi energy as measured relative to the bottom of the band and thus varies as the occupation changes. Depending on dimensionality, the ratio of the two capacitances is then given by
\begin{equation}
    \frac{C_{\rm G}}{C_{\rm Q}}=
    \begin{dcases}
      \frac{\hbar^2\pi^2\varepsilon\varepsilon_{0}w}{4m^{*}e^2D}n_{\rm 1D} & \text{for gate-wire system}, 
      \\
      \frac{\hbar^2\pi\varepsilon\varepsilon_{0}}{m^{*}e^2D} & \text{for gate-2DEG system}, \label{2DEG-gate}
    \end{dcases}  
\end{equation}
where $D$ is the distance from the wells to the surface, $w$ is the width of the wire, and $m^*$ is the electron effective mass in GaAs. Here, we assumed $n_{\rm 1D}=n_{\rm 2D}\cdot w$, with $n_{\rm 1D}$ and $n_{\rm 2D}$ as defined in equations \eqref{1D_density} and \eqref{2D_density}, respectively. Taking $w=50$\,nm, $D=85$\,nm, $m^*=0.067m_{\rm e}$, $\varepsilon=12$ and $n_{\rm 1D}=33$\,$\mu$m$^{-1}$, we get $C_{\rm G}/C_{\rm Q}=0.036$ and $C_{\rm G}/C_{\rm Q}=0.028$ for a 1D and 2D system, respectively. In both cases $C_{\rm Q}\gg C_{\rm G}$ and so the geometric contribution is expected to dominate in any capacitance measurement.

Our system consists of a GaAs/AlGaAs double-quantum-well heterostructure with translational invariance along the $x$- and $y$-directions, the wells being at roughly $70$ and $100$\,nm below the surface. A set of surface gates is used to define the quantum wires in the upper well. There is capacitive coupling between the wells and also between each well and the surface gates, which provide screening to the 2D electron gas.

We used the COMSOL Multiphysics 5.5 software package at \url{www.comsol.com} to simulate our device electrostatically (not self-consistently) and computed the electrical field and the potential distribution in the dielectrics given the known charge distributions in each well, one of which had wires defined by the surface gates (see Fig.\ \ref{fig:SF_FG}B and \ref{fig:SF_FG}C). Specifically, when solving Poisson equation we took $U(\textbf{r})$ as the potential induced by the gates and solved for it by taking $U(\textbf{r})=V_{\rm SG}$ at the gates and Neumann boundary conditions otherwise. In our simulation we accounted for the finite width of the 2D electron systems, though this made little difference to the results. Poisson's equation was solved using a finite-element grid, the number of nodes chosen so that the computation was free from finite-element effects. Finally, in our model, we ignored the effect of the ionised donor layers on both sides of the wells since they form a static layer of charge which is not affected by changes to gate voltage. The capacitance values (per unit area) obtained were $c=0.0047$\,Fm$^{-2}$ for the 2D injection region and $c=0.0124$\,Fm$^{-2}$ for the 1D wire region in the single-subband regime.

\subsubsection{2D-2D tunnelling signal}

\begin{figure*}[ht]
\centering
\includegraphics[width=0.95\textwidth]{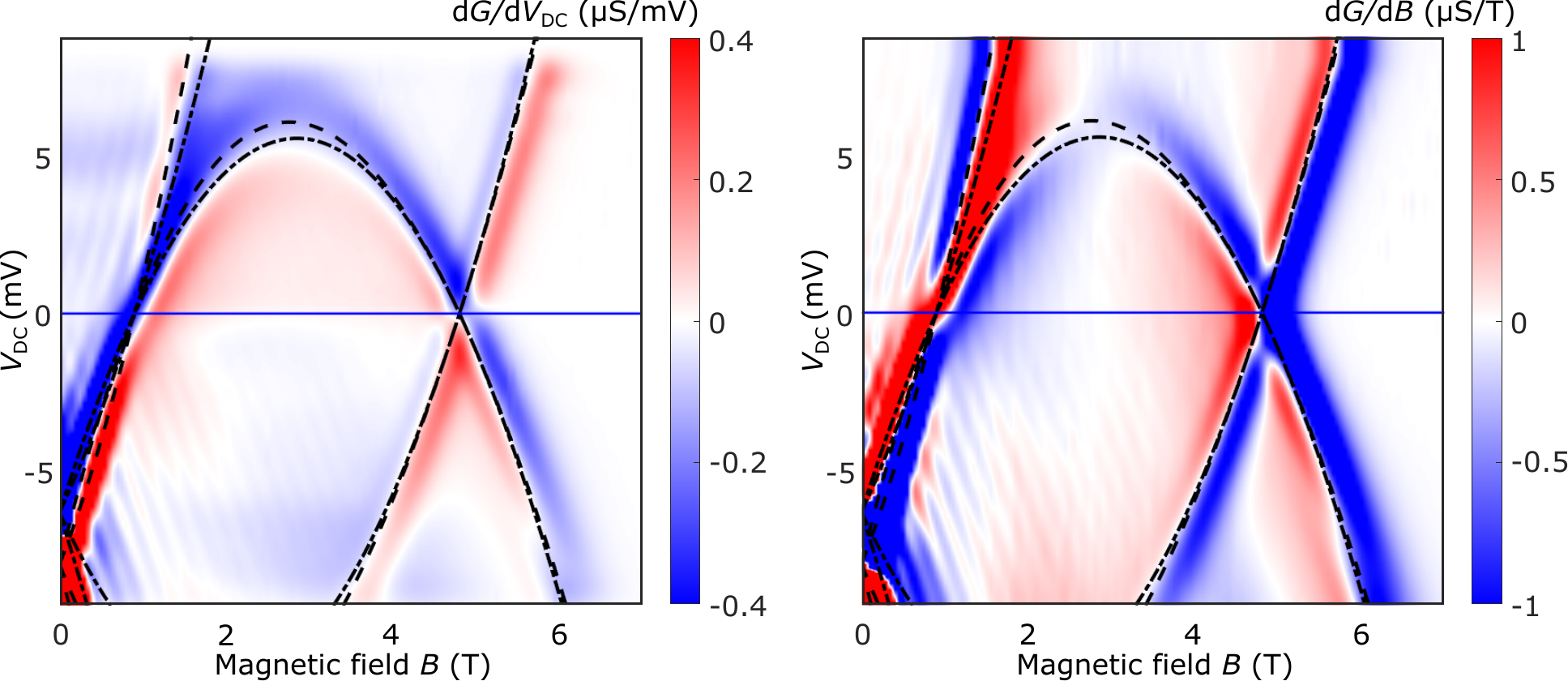}
\caption{\small \textbf{2D-2D Background.} Tunnelling conduction differentials $\textrm{d}G/\textrm{d}V_{\textrm{DC}}$ and $\textrm{d}G/\textrm{d}B$ vs magnetic field \textit{B} ($\propto$ momentum) and voltage $V_{\textrm{DC}}$ ($\propto$ energy $eV_{\textrm{DC}}$) for a 1.7\,$\mu$m long device, mapped at $V_{\textrm{WG}}=-0.57$\,V. At this voltage the wires are pinched off and do not conduct. Therefore, the spectrum observed arises solely from 2D-2D tunnelling taking place between the `parasitic' injection region and the bottom 2DEG, with the dashed and dash-dotted black curves corresponding to the capacitance-corrected and uncorrected resonances, respectively. This conductance is later subtracted from the tunnelling data obtained when the wires are conducting, allowing us to separate the `parasitic' from the 1D-2D tunnelling signal. Note that the quality of the fit with $d=$const.\ allows us to rule out a potential $d\equiv d(B)$ dependence due to the Lorentz repulsion between the wells, at least up to 7\,T, which is the highest field used in our experiment.}
\label{fig:SF_Background}
\end{figure*}

In order to subtract the influence of the `parasitic' injection region from the overall signal we map it beforehand by setting $V_{\textrm{WG}}$ negative enough that the wires do not conduct (see Fig.\ \ref{fig:SF_Background}, below the bottom subband). The parameters of interest when fitting in this regime are, respectively, the capacitances (per unit area) of both the upper and lower wells, $c_{\textrm{UW}}^{\textrm{2D}}$ and $c_{\textrm{LW}}^{\textrm{2D}}$, the effective masses of electrons in top and bottom wells, $m_{\textrm{UW}}$ and $m_{\textrm{LW}}$, the separation between the wells, $d$, and the zero-bias field intercepts, $B_{-}^{\textrm{p}}$ and $B_{+}^{\textrm{p}}$. Given that this region is relatively wide ($\sim 0.45\mu$m), it can be safely treated as a 2D system in both wells. 

We found that good results were obtained when setting both $m_{\textrm{UW}}$ and $m_{\textrm{LW}}$ to be equal $0.93m_{\textrm{b}}$, where $m_{\textrm{b}}=0.067m_{\textrm{e}}$ is the electron band mass in GaAs. We note that $m_{\textrm{2D}}^{\textrm{*}}=0.93m_{\rm b}=0.062m_{\textrm{e}}$ is in very good agreement with independent work carried out in 2D systems very similar to ours \cite{coleridge_effective_1996,hayne_exchange_1992,hatke_evidence_2013,tan_measurements_2005,kwon_quantum_1994}. From the MBE growth data we also know that $d\approx32$\,nm and, even though it is reasonable to expect deviations from this value due to monolayer fluctuations, these should not exceed a few nanometres. Finally, the zero-bias crossing points can be easily determined by visual inspection of the data while the capacitive coupling was estimated by simulating it under an electrostatic framework following the approach described in the previous section. We found that the best results were obtained with $d=31$\,nm, $c_\textrm{UW}^\textrm{2D}=0.0047$\,Fm$^{-2}$ and $c_\textrm{LW}^\textrm{2D}=0.0033$\,Fm$^{-2}$, in very good agreement with the expected values. Owing to extra coupling to the surface gates, the calculated and observed capacitances of the upper well are approximately $\sim 100/70=1.4$ that of the lower well.

It is worth noting also that, particularly at high fields where the sign of $k$ changes while tunnelling, one could expect the Lorentz force to act in such a way as to force carriers in each well further apart. This, in practice, would translate in having $d\equiv d(B)$. However, as can be seen from Fig.\ \ref{fig:SF_Background}, no such correction is needed up to at least 7\,T and so we can safely rule it out for the rest of our analysis.

\subsubsection{1D-2D tunnelling signal}

In the previous section we discussed how to constrain $c_{\textrm{UW}}^{\textrm{2D}}$, $c_\textrm{LW}^{\textrm{2D}}$, $m_{\textrm{2D}}^*$, and $d$ from the background data, \emph{i.e.} using the conductance maps with the 1D wires past pinch-off. Since the observed modulation by the wire gate to the injection region is very small however, as can be seen by noting that the density under the parasitic and the wire regions closely match here (see Fig.\ \ref{fig:SF_FG}C), it is reasonable to expect the same values to apply when fitting to the 2D signal even as the 1D channels are now conducting. 

We map the 1D subbands as shown in Fig.\ \ref{fig:SF_FG}A. Under the current geometry, most devices usually display between three to four subbands below the 2D band. By setting appropriate values for $V_{\textrm{WG}}$, we are able to map the dispersion of the system at different subband occupancy values (see an example in Fig.\ 4A in the main text). For now, we will restrict ourselves to the single-subband regime. From Fig.\ \ref{fig:SF_FG}, both the Fermi energies as well as the 1D (2D) densities in the wire (`parasitic') region can be determined by using equations \eqref{layer_fermi_vectors}, \eqref{1D_density} and \eqref{2D_density}.  

\begin{figure*}[h!]
\centering
\includegraphics[width=0.80\textwidth]{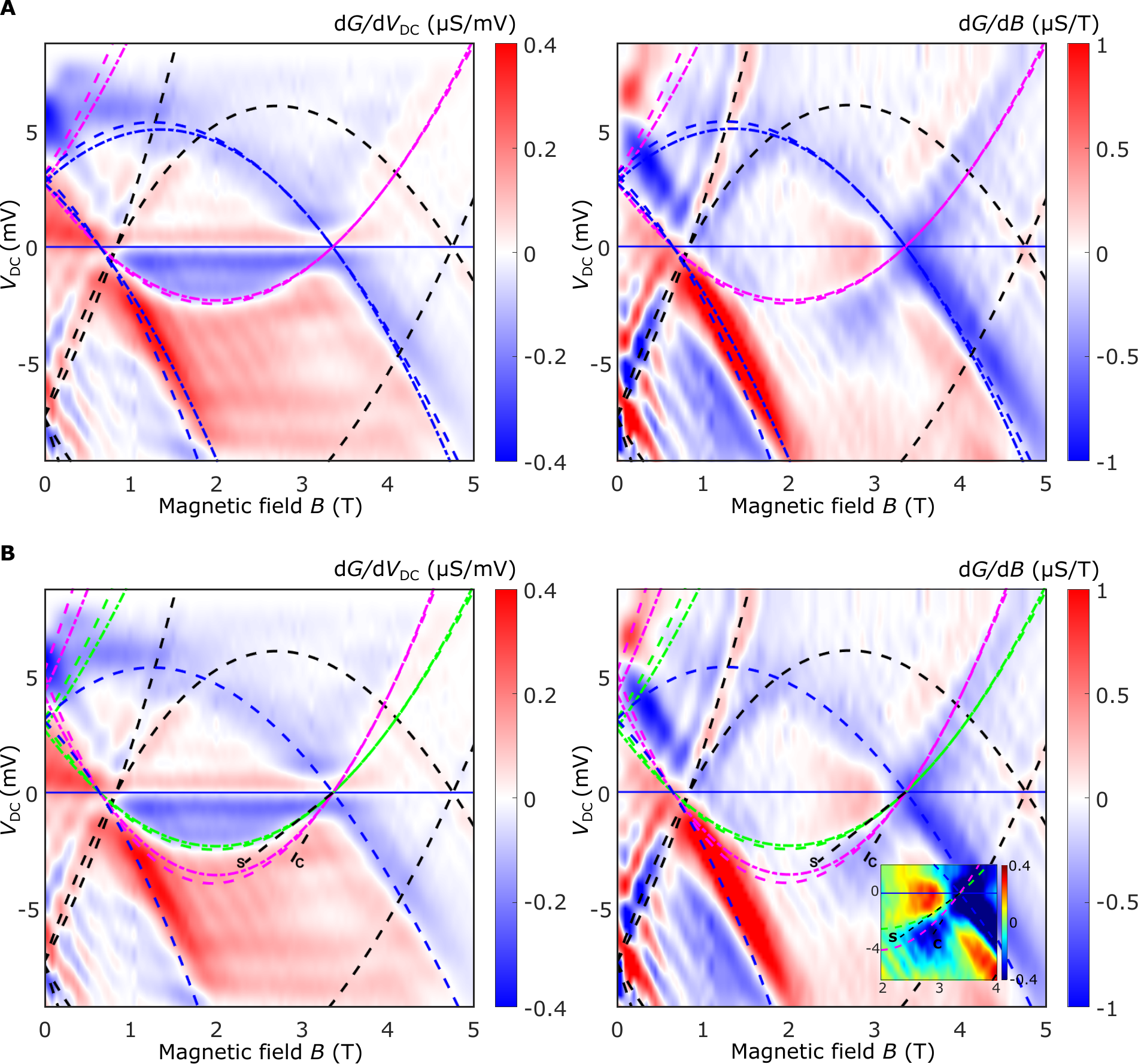}
\caption{\small \textbf{Single-subband regime: One Fermi Sea.} Tunnelling-conductance differentials $\textrm{d}G/\textrm{d}V_{\textrm{DC}}$ and $\textrm{d}G/\textrm{d}B$ as a function of the DC-bias $V_{\textrm{DC}}$ and the in-plane magnetic field $B$, for a 1.7\,$\mu$m device, mapped at $V_{\textrm{WG}}=-0.515$\,V. (\textbf{A}) Model 1, with a single plasmon type, where $m_{\textrm{1D}}^{*}=m_{\textrm{2D}}^{*}=0.93\,m_{\textrm{b}}$. The dashed black curves indicate the location of the subtracted 2D-2D `parasitic' tunnelling signal mapped in Fig.\ \ref{fig:SF_Background}. The dashed and dash-dotted magenta curves mark the capacitance corrected and uncorrected resonances arising from the tunnelling between the LW ground states and the UW wire region. They reveal the dispersion of the elementary excitations in the 1D UW wire region. Similarly, the dashed and dash-dotted blue curves mark the location of the resonances resulting from the reverse tunnelling process, between the UW ground states and the LW, revealing the dispersion of the 2D LW. (\textbf{B}) Same as in (\textbf{A}) but now fitted using Model 2, where two types of excitations, spin and charge, are allowed. The dashed black and blue lines represent the same dispersions as before. The green curves correspond to a spinon mode with $m_{\textrm{s}}=m_{\textrm{2D}}^{*}$. The magenta lines, on the other hand, mark a holon mode with $m_{\textrm{c}}<m_{\textrm{2D}}^{*}$. Inset: Spin-charge separation (dashed black `S' and `C' lines) near the +$k_{\rm F}$ point at low bias.}

\label{fig:SF_M1_M2}
\end{figure*}

The procedure followed when fitting the tunnelling signal arising from the wire region is analogous to what was done for the `parasitic' injection region. Since $c_{\textrm{UW}}^{\textrm{2D}}$, $c_{\textrm{LW}}^{\textrm{2D}}$ and $d$ have already been obtained, the only parameters left to be determined are $B_-^{\textrm{w}}$ and $B_+^{\textrm{w}}$ (the zero-bias field intersects for the 1D signal, obtained similarly by visual inspection of the tunnelling maps), $m_{\textrm{UW}}=m_{\textrm{1D}}^{\textrm{*}}$ (the effective mass of the 1D electrons in the upper well), and $c_{\textrm{UW}}^{\textrm{1D}}$ and $c_{\textrm{LW}}^{\rm 1D}$ (the capacitances per unit area in both wells). For the latter the values found, $c_{\textrm{UW}}^{\textrm{1D}}=c_{\textrm{LW}}^{\rm 1D}=0.0026$\,Fm$^{-2}$, were significantly lower that those predicted in COMSOL. Using the larger values for capacitance completely distorts the dispersions and cannot be accounted for by errors in other parameters as the constraints are strong. 
We interpret the observed deviations as a significantly stronger effect of interactions in 1D that push the band down, reducing the rate of filling the wire, $dn_{\rm 2D}/dE_{\rm F}$, or in other words, increasing the energy required to add an extra electron. This results in a decrease in the quantum capacitance $C_{\rm Q}$, therefore making the total capacitance $C$ no longer solely determined by the geometric contributions as was done in our simulation, see \eqref{capacitances}.

The effective mass of electrons in the lower well in the 1D-2D regime was found to be slightly higher than in the 2D-2D scenario ($\simeq 0.067m_{\textrm{e}}$ vs $\simeq 0.062m_{\textrm{e}}$) that we attribute to the different amount of interactions, which electrons in the Fermi liquid of the bottom layer experience when the wire-gate depletes significantly the upper layer. In other words, reduction in screening by the upper layer changes interactions and hence $m_{\textrm{2D}}^*$.

\textbf{Model 1: Single Fermi Sea and Single Plasmon} Let us first assume that there is a single, uniquely determined 1D effective mass, which matches the 2D non-interacting effective mass, $m_{\textrm{1D}}^{\textrm{*}}=m_{\textrm{2D}}^{\textrm{*}}$. From Fig.\ \ref{fig:SF_M1_M2}A it can be seen then that while the 1D mode is nicely fitted in the hole sector ($V_{\textrm{DC}}<0$), the same does not happen in the particle sector ($V_{\textrm{DC}}>0$), particularly at very high biases, even when correcting for capacitance. We therefore conclude that the 1D excitation cannot be fully captured by a single 1D effective mass.

\textbf{Model 2: Single Fermi Sea and Two Plasmons} We now assume that there are both spin- and charge-type excitations in our 1D system. We note that we have previously observed spin-charge separation in similar devices \cite{Jompol09}. Under our configuration, we also know that the 1D excitations in the hole sector correspond to a spinon mode, while that in the particle sector to a holon mode \cite{Moreno16}. Since each mode corresponds to a different kind of excitations, it is reasonable to expect different effective masses for each, which we label as $m_{\textrm{s}}$ and $m_{\textrm{c}}$ respectively. As before, it can be shown (see Fig.\ \ref{fig:SF_M1_M2}B) that while $m_{\textrm{s}}=m_{\textrm{2D}}^{\textrm{*}}$ provides a good match to the data, even when taking $m_{\textrm{c}}<m_{\textrm{2D}}^{\textrm{*}}$ one still fails to fully capture the observed behaviour for this mode. This is particularly noticeable when taking the $\textrm{d}G/\textrm{d}B$ differential. We also note that any fit to the holon branch in the particle sector must also catch the low-energy charge mode, that is, the holon branch in the hole sector, when extended back to it, which is clearly not the case.

\textbf{Model 3: Two Fermi Seas and Two Plasmons} Based on the previous analyses we will now consider the spectroscopic predictions of the 1D Fermi-Hubbard model (see details in the Supplementary Text `Spin and charge excitations' above and Fig.\ \ref{fig:SF_hs_excitations}C), where we now assume not only different effective masses $m_{\textrm{s}}$ and $m_{\textrm{c}}$ but also different densities, and therefore different Fermi momenta for both the spinon and the holon modes. We start from the assumption that the dispersion of both modes, $f$, is parabolic, before correcting for capacitance. That gives us
\begin{equation}
    f_1(k)=\frac{\hbar^2k^2}{2m_{\textrm{s}}}-E_{\textrm{F}}^{\textrm{s}}=\frac{\hbar^2}{2m_{\textrm{s}}}(k^2-k_{\textrm{F}}^2)
\end{equation}
and
\begin{equation}
    f_2(k)=\frac{\hbar^2(k+k_{\textrm{F}})^2}{2m_{\textrm{c}}}-E_{\textrm{F}}^\textrm{c}=\frac{\hbar^2}{2m_{\textrm{c}}}\left[(k+k_{\textrm{F}})^2-4k_{\textrm{F}}^2\right]
\end{equation}
where $E_{\textrm{F}}^{\textrm{s}}$, $m_{\textrm{s}}$ and $E_{\textrm{F}}^{\textrm{c}}$, $m_{\textrm{c}}$ are the respective Fermi energies and the renormalised masses, with $k_{\rm F}\equiv k_{\rm F,UW}$ for simplicity. The charge mode, forming between $-3k_{\textrm{F}}$ and $+k_{\textrm{F}}$, has a density of states half that of the spinon excitation, between $\pm k_{\textrm{F}}$. An analogous expression can also be obtained for the holon branch between $-k_{\textrm{F}}$ and $+3k_{\textrm{F}}$. Since $v=\textrm{d}\omega/\textrm{d}k=\hbar^{-1}\textrm{d}E/\textrm{d}k$, we get
\begin{equation}
    v^{\textrm{s}}=\left. \frac{1}{\hbar}\frac{\textrm{d}f_1}{\textrm{d}k}\right|_{k=k_{\textrm{F}}}=\left. \frac{\hbar k}{m_{\textrm{s}}}\right|_{k=k_{\textrm{F}}}=\frac{\hbar k_{\textrm{F}}}{m_{\textrm{s}}}
\end{equation}
and
\begin{equation}
    v^{\textrm{c}}=\left. \frac{1}{\hbar}\frac{\textrm{d}f_2}{\textrm{d}k}\right|_{k=k_{\textrm{F}}}=\left. \frac{\hbar (k+k_{\textrm{F}})}{m_{\textrm{c}}}\right|_{k=k_{\textrm{F}}}=\frac{2\hbar k_{\textrm{F}}}{m_{\textrm{c}}},
\end{equation}
therefore arriving at
\begin{equation}\label{S13}
    \frac{v^{\textrm{c}}}{v^{\textrm{s}}}=\frac{K_{\textrm{s}}}{K_{\textrm{c}}}=\frac{2m_{\textrm{s}}}{m_{\textrm{c}}}.
\end{equation}

\begin{figure*}[ht!]
\centering
\includegraphics[width=0.9\textwidth]{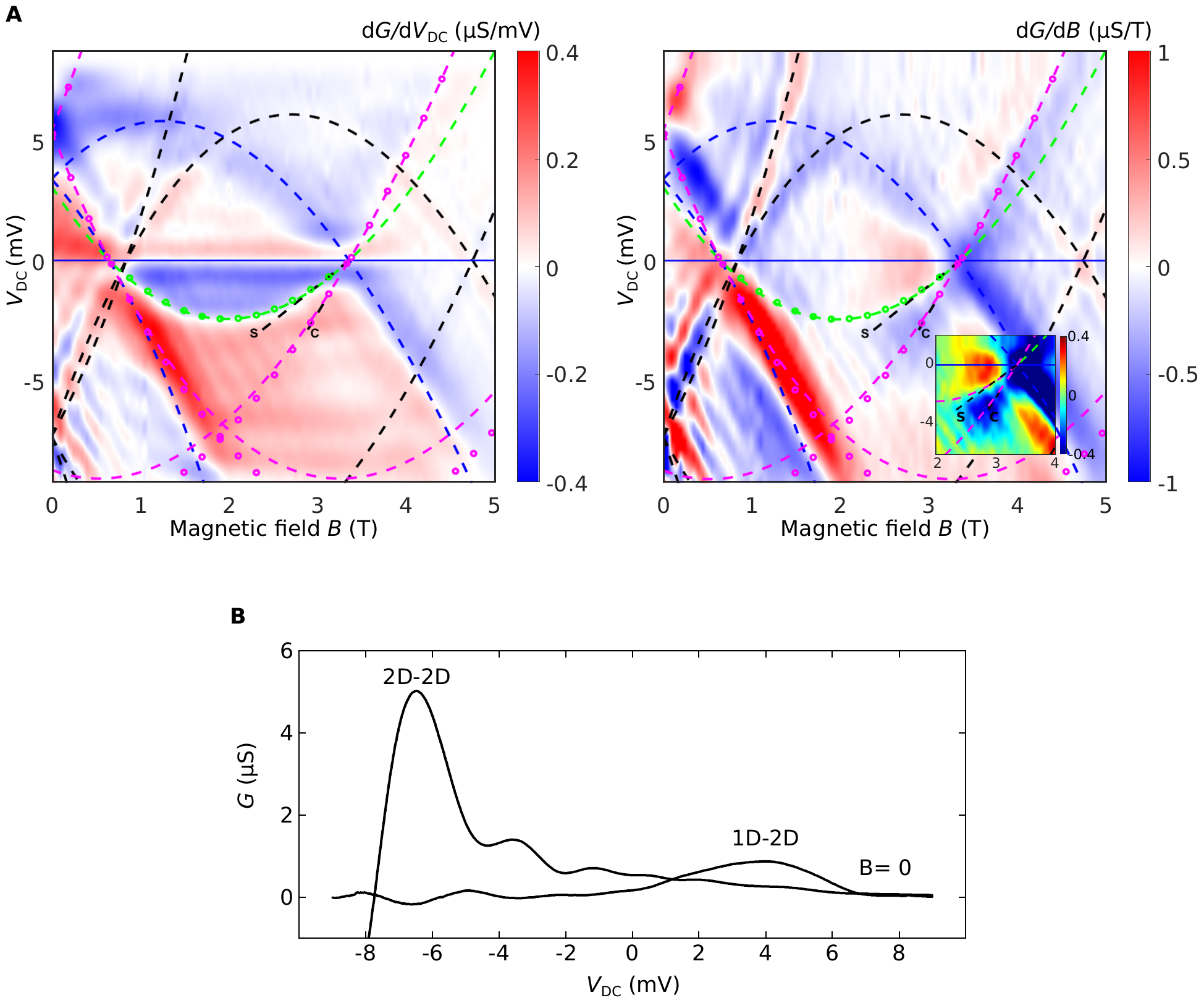}
\caption{\small \textbf{Single-subband regime: Two Fermi Seas and Fermi-Hubbard Model.} (\textbf{A}) Tunnelling-conductance differentials $\textrm{d}G/\textrm{d}V_{\textrm{DC}}$ and $\textrm{d}G/\textrm{d}B$ as a function of the DC-bias $V_{\textrm{DC}}$ and the in-plane magnetic field $B$, for a 1.7\,$\mu$m device, mapped at $V_{\textrm{WG}}=-0.515$\,V and fitted using the model of two Fermi seas, see text. Dashed black and blue curves have the same meanings as before, see Fig.\ \ref{fig:SF_M1_M2}. The dashed green and magenta lines mark the 1D dispersions for spinons and holons, respectively. The open-circle curves mark the corresponding solutions of the Fermi-Hubbard model for $N=54$. Inset: Spin-charge separation (dashed black `S' and `C' lines) near the +$k_\textrm{F}$ point at low-bias. (\textbf{B}) Conductance $G$ line-cuts at $B=0$, showing the broadened 1D-2D tunnelling resonance peak, and the equivalent 2D-2D background-calibration trace with the wires pinched off. The broad 1D-2D peak comes from the overlap of the dashed blue and dashed magenta resonances as described in \textbf{A}, see section 1.6.4 for more details.}
\label{fig:SF_M3}
\end{figure*}


Equation \eqref{S13} allows us to relate the charge-to-spin velocity ratio with the phenomenological Luttinger parameters $K_{\textrm{c,s}}$, which account for the renormalisation of the effective masses $m_{\textrm{s}}$ and $m_{\textrm{c}}$ due to the 1D confinement. We have $m_{\textrm{s}}=m_{\textrm{b}}K_{\textrm{s}}$ and $m_{\textrm{c}}=m_{\textrm{b}}K_{\textrm{c}}$, with $K_{\textrm{c}}\neq K_{\textrm{s}}$. The ratio $K_{\textrm{c}}/K_{\textrm{s}}$ is a good estimate of the interaction strength. Note also that the extra factor of two in the third term of \eqref{S13} arises from the assumption of different densities of states of the collective modes in each Fermi sea, or in other words because $k_{\textrm{F}}$ is twice as large for holons as it is for spinons. For repulsive interactions, since $K_{\textrm{s}}>1$ and $K_{\textrm{c}}<1$, we have $m_{\textrm{c}}<2m_{\textrm{s}}$.

In Fig.\ \ref{fig:SF_M3}A we fit the data using the model with two Fermi seas. The dashed curves were obtained using the parabolic model described in this section while the open-circle lines correspond to solutions of the Fermi-Hubbard model. Both models are in very good agreement across all momentum and energy range experimentally probed. Note how, unlike before, the holon mode emanating from $+k_{\textrm{F}}$ is now fully captured both in the hole and particle sectors. The line shapes at zero magnetic field for various tunnelling processes are also shown in Fig.\ \ref{fig:SF_M3}B. The tunnelling resonance peak at large negative bias (black dashed line in Fig.\ \ref{fig:SF_M3}A) corresponds to 2D-2D `parasitic' tunnelling, while at positive bias the main contribution to conductance comes from both 1D and 2D processes. Unlike the 2D-2D peak however, some significant amount of broadening can be observed. We interpret this as the superposition of both the 2D (dashed blue) and 1D (dashed magenta) dispersions, as predicted by our model, see the next section for discussion. Most importantly, however, this feature is completely missed by previous models, see Fig.\ \ref{fig:SF_M1_M2}.

\subsubsection{Tunnelling resonances at zero field}

\begin{figure}[ht!]
  \includegraphics[width=0.9\textwidth]{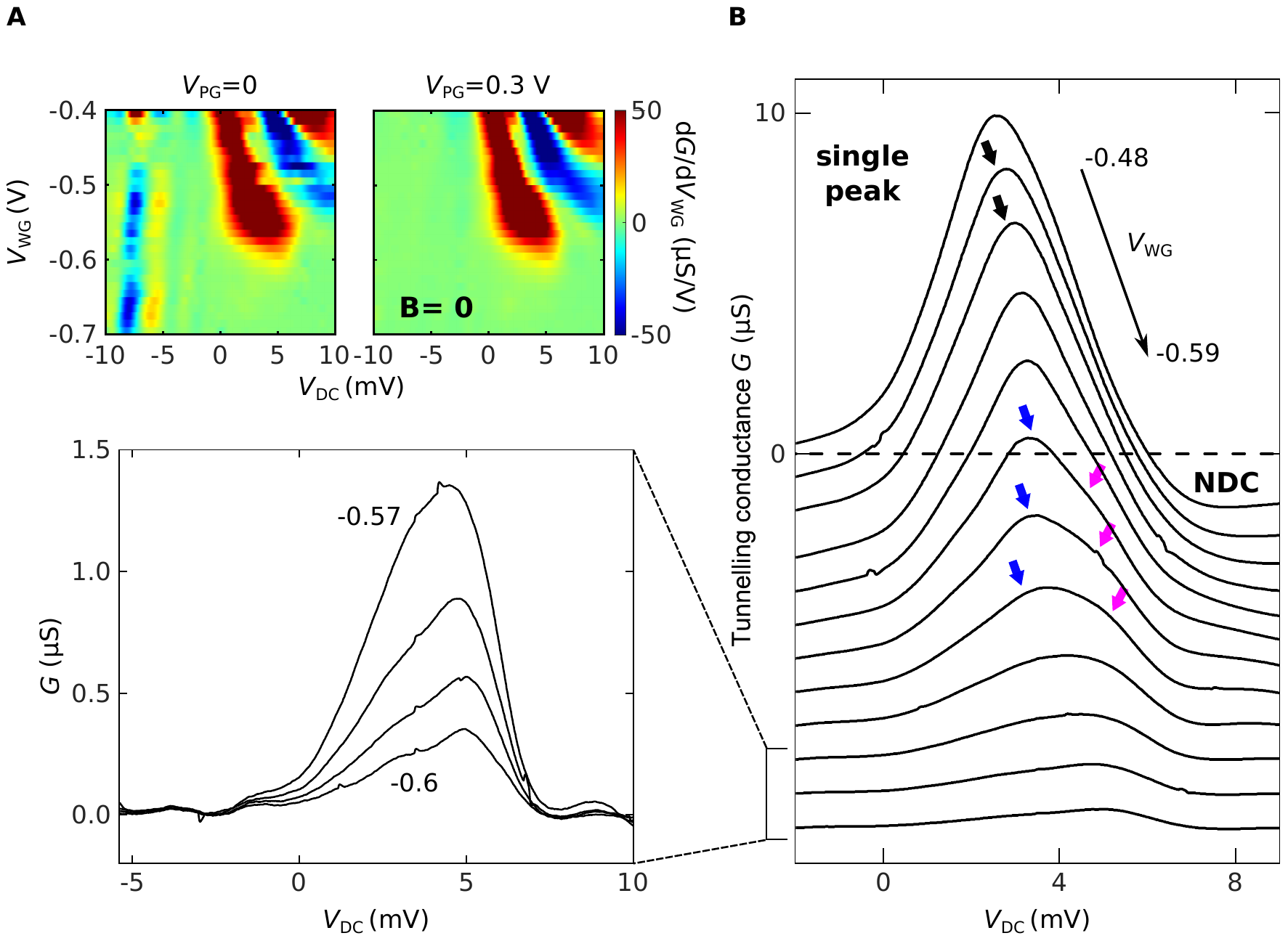}
  \centering
  \caption{\small \textbf{Tunnelling processes at zero field.} (\textbf{A}) $\textrm{d}G/\textrm{d}V_\textrm{WG}$ vs $V_\textrm{WG}$ and $V_\textrm{DC}$, at $B=0$\, for both $V_\textrm{PG}=0$ and $V_\textrm{PG}=0.3$\,V (the latter moves the fringes arising in the `parasitic' region out of the range of the plot). The conductance peak seen at $V_\textrm{DC}>$0 disappears at $V_\textrm{WG}=-0.6$\,V, the same value at which the wires pinch off in Fig.\ 1C; (\textbf{B}) Horizontal line-cuts of the conductance data shown in \textbf{A} for positive $V_\textrm{PG}$. Here, the signal can be seen to slowly broaden as $V_\textrm{WG}$ is made progressively more negative (\emph{i.e.}, by increasing the level of confinement in the wires and hence decreasing the density), eventually separating into two distinct features, with a similar spacing to that of the 2D and 1D holon modes in the particle sector. Every curve, except for $V_\textrm{WG}=-0.48\,$V, has been offset for clarity. The left-hand panel shows, without any offsets, the line-cuts for $V_\textrm{WG}\leq-0.57$\,V (\emph{i.e.}, as the 1D channels pinch off). 
  }
  \label{fig:SF_Zero_field}
\end{figure}

Besides the full mapped dispersions, we can also inspect the various tunnelling processes seen at zero magnetic field (see Fig.\ \ref{fig:SF_Zero_field}A for a 5\,$\mu$m device) to further constrain our model. The two panels show $\textrm{d}G/\textrm{d}V_\textrm{WG}$ vs wire-gate voltage $V_\textrm{WG}$ and DC bias $V_\textrm{DC}$, for two values of the voltage $V_\textrm{PG}$ on a gate that allows us to move the `parasitic' features around by changing the density in the `parasitic' region. All features at negative DC bias move away for positive $V_\textrm{PG}$, indicating they are not related to the 1D wires themselves. They correspond to the 2D-2D `parasitic' tunnelling discussed in section 1.6.2. On the other hand, at positive $V_{\textrm{DC}}$, the main contribution comes from the superposition of both 1D and 2D tunnelling happening in the wire region, which was discussed in section 1.6.3. Note that, not only are these features unaffected by $V_\textrm{PG}$, but they also disappear around $V_\textrm{WG}=-0.6$\,V, the same value at which the wires are known to pinch off, see Fig.\ 1C. Fig.\ \ref{fig:SF_Zero_field}B shows the line profiles of the data shown in Fig.\ \ref{fig:SF_Zero_field}A for positive $V_\textrm{PG}$. As can be seen, the tunnelling signal corresponding to the wires slowly broadens as the confinement is made progressively stronger (by making $V_\textrm{WG}$ more negative, hence reducing the density).

At $B=0$ there should be a peak in current as a function of $V_\textrm{DC}$ as the concentric spectral functions of the 1D and 2D systems move in and out of alignment. The conductance will therefore have a peak and a trough (negative differential conductance, NDC), like that shown at small wire-gate voltages in Fig.~\ref{fig:SF_Zero_field}B. Higher 1D subbands and the small overlap of a large number of states in the two systems (because at $B=0$ their spectral functions are not offset in momentum), enhance the tunnelling above the peak, reducing the depth of the trough and moving the peak position towards higher bias. Nevertheless, as the wires are squeezed, from $V_\textrm{WG}=-0.53$\,V onwards, first a kink, then two well-defined features can be seen, unaffected by $V_\textrm{PG}$ and therefore not arising from the parasitic region. While this extra feature could indicate tunnelling into an empty second 1D subband, there is no sign of this anywhere else in our full energy--momentum maps such as that in Fig.~2C. Also, the splitting is much too great to be explained by different effective masses of the 2D Fermi-liquid quasi-particles and the 1D spinon modes. Note also that, for the latter case, the spinon mode would be equivalent to that coming from the $+k_\textrm{F}$ point, which we observe to be absent in the particle sector, and is also therefore not expected here below $-k_\textrm{F}$.

The observed feature is separated from the main peak by about the same bias as the 2D (dashed blue) and 1D holon (dashed magenta) branches emanating from the $-k_\textrm{F}$ point in the particle sector at $B=0$. For each wire-gate voltage, we find good agreement between these zero-field features and the position of the 2D and 1D holon spectral modes as predicted from the full dispersion maps. This is completely missed by Models 1 and 2 previously discussed, and can only be successfully accounted for if two Fermi seas are assumed.

\subsubsection{Multiple-subband occupancy}

Every time a new subband starts being occupied, another four parameters ($B_-^{\textrm{w}}$, $B_+^{\textrm{w}}$, $m_{\textrm{s}}$ and $m_{\textrm{c}}$) need to be added. The capacitances $c_{\rm UW}^{\rm 1D}$ and $c_{\rm LW}^{\rm 1D}$ also need to be updated, with their values now lying somewhere in between the single-subband 1D-2D and the 2D-2D regimes. This leads to, for example, a total of 6+4+4+4=18 parameters when in the three subband regime. 

\begin{figure*}[ht!]
\centering
\includegraphics[width=0.9\textwidth]{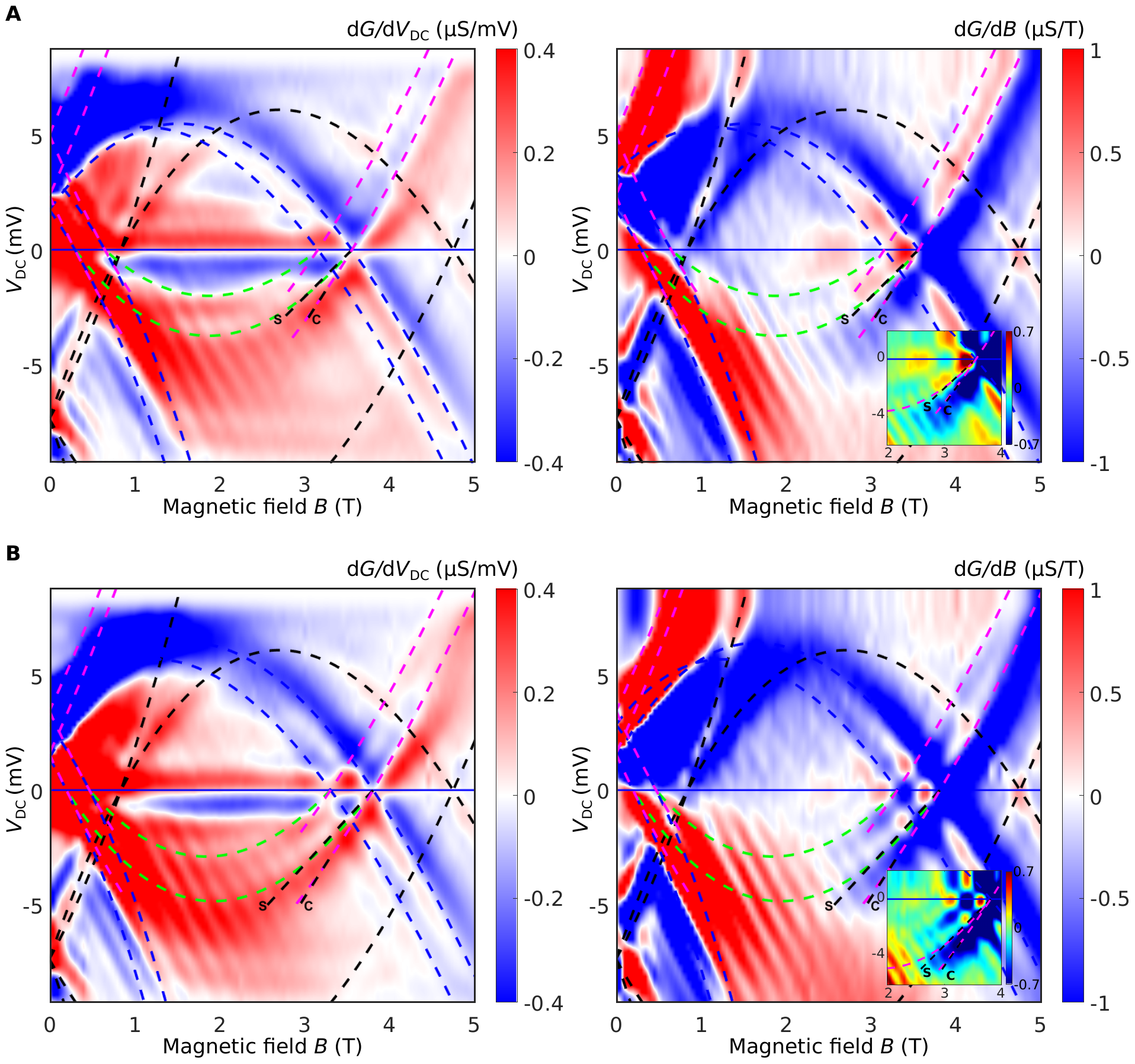}
\caption{\small \textbf{Multiple-subband regime.} Tunnelling conductance differentials $\textrm{d}G/\textrm{d}V_{\textrm{DC}}$ and $\textrm{d}G/\textrm{d}B$ plotted as a function of the DC-bias $V_{\textrm{DC}}$ and the in-plane magnetic field $B$ for a 1.7\,$\mu$m device, mapped with (\textbf{A}) two ($V_{\textrm{WG}}=-0.455$\,V) and (\textbf{B}) three ($V_{\textrm{WG}}=-0.405$\,V) occupied subbands, respectively. The dashed coloured curves have the same meaning as before (see Fig.\ \ref{fig:SF_M1_M2} and \ref{fig:SF_M3}). Insets show spin-charge separation near the +$k_{\textrm{F}}$ point for the bottom-most occupied subband.} 
\label{fig:SF_Multiple_subbands}
\end{figure*}

We have restricted our analysis, where applicable, to the bottom two subbands even when more are occupied, as it becomes progressively more difficult to accurately and reliably analyse the data, see Fig.\ \ref{fig:SF_Multiple_subbands} and Fig.\ 4 in the main text. 
One of the difficulties is that, as the number of subbands is increased, it becomes visually harder to extract $B_-^{\textrm{w}}$ for most subbands. We know however that a high degree of symmetry exists between all 1D subbands (see Fig.\ \ref{fig:SF_FG}A). Since the 1D channels have approximately parabolic confinement potentials, the subband dispersions should approximately follow that of a harmonic oscillator. Therefore, we have
\begin{equation}
    \frac{B_{+,2}+B_{-,2}}{2}=\frac{B_{+,1}+B_{-,1}}{2},
\label{subband_symmetry}
\end{equation}
where $B_{\pm,n}$ denote the crossing points between the nth-subband and the $B$ axis at $V_{\textrm{DC}}=0$. Similarly, one can look at the crossing point $V_{\rm 0,n}$ between the $n$th-subband and the $V_\textrm{DC}$ axis at $B= 0$, in which case we obtain
\begin{equation}
eV_{0,n}=\pm\frac{\hbar^2}{2m^\star}(k_{\textrm{F,UW,}n}^2-k_\textrm{F,LW}^2)\Longrightarrow V_{0,n}=\pm\frac{ed^2}{2m^\star}B_{+,n}B_{-,n}.
\label{zero_field_crossing}
\end{equation}
Here, $k_{\textrm{F,UW,}n}$ refers to the Fermi wavevector of the spinon mode of the $n$th subband while $\pm$ labels the lower and upper layers, respectively. The second equality is obtained by converting $k_\textrm{F}$s into $B_{\pm,n}$ using (\ref{layer_fermi_vectors}). Combining both (\ref{subband_symmetry}) and (\ref{zero_field_crossing}) we arrive at
\begin{equation}
\begin{split}
    (B_{+,2}-B_{-,2})^2=(B_{+,2}+ & B_{-,2})^2-4B_{+,2}B_{-,2} \\ & \Longrightarrow \frac{B_{+,2}-B_{-,2}}{2}=\sqrt{\bigg(\frac{B_{+,1}+B_{-,1}}{2}\bigg)^2\pm V_{0,2}\frac{2m}{ed^2}} .
\end{split}
\label{multi_subband}
\end{equation}
Equation (\ref{multi_subband}) can be used to determine the position of the second subband by varying $V_{0,2}$ and plotting the dispersions with the resultant $B_{\pm,2}$ until the best match is obtained. Since $B_{+,2}$ is usually well-observed in the data, this effectively imposes a very strong constraint on $B_{-,2}$.

Another difficulty of the multiple-occupancy regime has to do with extracting the holon and spinon masses, $m_\textrm{c}$ and $m_\textrm{s}$. The first is usually obtained by looking at the holon branch emerging from $+k_\textrm{F}$ at high bias. However, as more subbands come into play these become harder to resolve separately. Similarly to what was done in the single-subband regime, we can obtain a second constraint on $m_\textrm{c}$ by looking at the peak broadening at $B=0$. On the other hand, $m_\textrm{s}$ is extracted by matching to the spinon-mode dispersion in the hole sector. This is usually clearly observed for at least the bottom two subbands. Nevertheless, since the spin and charge velocities can only be obtained for the bottom-most occupied subband, this ratio cannot be compared to the independently obtained mass ratios, as was done when only one subband was occupied. Overall, this translates into a slight increase of the error with higher subband occupancy, as seen from Fig.\ 2E and 4 in the main text.

\bibliography{references.bib}
\bibliographystyle{ScienceAdvances}